\documentclass[twocolumn,nofootinbib,longbibliography]{revtex4-1}

\usepackage[english]{babel}
\usepackage{appendix}
\usepackage{natbib}

\usepackage{amsmath}
\usepackage{mathrsfs}
\usepackage{amssymb}
\usepackage{amsfonts}
\usepackage{dsfont}
\usepackage{IEEEtrantools}
\usepackage{bm}
\usepackage{amsthm}

\usepackage[breaklinks=true]{hyperref}
\usepackage[inline]{enumitem}

\makeatletter
\usepackage[threshold=1, autopunct, autostyle]{csquotes}
\@ifclassloaded{beamer}
  {}
  {
    \bibpunct{[}{]}{,}{n}{}{,}

  }
\makeatother

\DeclareMathOperator{\Prb}{P}

\newcommand{\cProb}[2]{\Prb(#1\!\mid\!#2)}

\newcommand{\Bt}{{\tt t}}
\newcommand{\Bf}{{\tt f}}

\DeclareMathOperator{\End}{End}

\usepackage{pgfplots}
\usepackage{tikz}
\usetikzlibrary{shapes,calc,positioning,arrows.meta}

\pgfdeclarelayer{bg}    \pgfsetlayers{bg,main}

\tikzset{colorbox/.style={thick, rounded corners=2pt, text height=1.7ex,text depth=.25ex, draw=#1!70!black, fill=#1!30}}
\tikzset{colorboxS/.style={thick, rounded corners=2pt, text height=1.2ex,text depth=.25ex, draw=#1!70!black, fill=#1!30, font=\footnotesize}}
\tikzset{colorboxXS/.style={thick, rounded corners=2pt, text height=0.7ex,text depth=.10ex, draw=#1!70!black, fill=#1!30, font=\tiny}}
\tikzset{roundedbox/.style={thick, rounded corners=2pt, text height=1.7ex,text depth=.25ex, draw=black}}
\tikzset{CBedgy/.style={thick, text height=1.7ex,text depth=.25ex, draw=#1!70!black, fill=#1!30, font=\ttfamily}}
\tikzset{colorelement/.style={thick, rounded corners=2pt, draw=#1!70!black, fill=#1!30}}
\tikzset{quote/.style={thick, draw=black!70, rounded corners=2pt, font=\footnotesize, text width=#1, text opacity=1, opacity=.8, fill=white},
  quote/.default=4.5cm}
\tikzset{quoteS/.style={thick, draw=black!70, rounded corners=2pt, font=\scriptsize, text width=#1, text opacity=1, opacity=.8, fill=white},
  quoteS/.default=4.5cm}
\tikzset{quoteXS/.style={thick, draw=black!70, rounded corners=2pt, font=\tiny, text width=#1, text opacity=1, opacity=.8, fill=white},
  quoteXS/.default=4.5cm}
\tikzset{quoteNB/.style={font=\footnotesize, text width=#1, text opacity=1, opacity=.8, fill=white},
  quoteNB/.default=4.5cm}
\tikzset{comment/.style={thick, draw=black!70, rounded corners=2pt, font=\scriptsize\itshape, text width=#1, text opacity=1, opacity=.8, fill=white},
  comment/.default=4.5cm}
\tikzset{commentS/.style={thick, draw=black!70, rounded corners=2pt, font=\tiny\itshape, text width=#1, text opacity=1, opacity=.8, fill=white},
  commentS/.default=4.5cm}
\tikzset{commentVW/.style={thick, draw=black!70, rounded corners=2pt, font=\scriptsize\itshape, text opacity=1, opacity=.8, fill=white}}
\tikzset{commentSVW/.style={thick, text height=0.7ex,text depth=.10ex, draw=black!70, rounded corners=2pt, font=\tiny\itshape, text opacity=1, opacity=.8, fill=white}}
\tikzset{conn/.style={thick, shorten <=#1, shorten >=#1}}
\tikzset{tconn/.style={shorten <=#1, shorten >=#1}}
\tikzset{gconn/.style={thick, shorten <=#1, shorten >=#1, draw=gray!80}}
\tikzset{arr_node/.style={pos=0.5,above,font=\scriptsize, sloped}}
\tikzset{ctag/.style={thick, dashed, rounded corners=2pt, text height=1.7ex,text depth=.25ex, draw=#1!70!black, fill=#1!30, font=\ttfamily}}

\tikzset{ist/.style={thick, shorten <=4pt, shorten >=4pt, arrows = {-Bracket[reversed,round]}}}
\tikzset{istgleich/.style={thick, shorten <=4pt, shorten >=4pt, arrows = {Bracket[reversed,round]-Bracket[reversed,round]}}}
\tikzset{nicht/.style={thick, dashed, shorten <=4pt, shorten >=4pt, arrows = {Bracket[round]-}}}

\definecolor{yorange}{HTML}{ff8c00}

\newcommand\blArrow{}
\def\blArrow[#1](#2);
  { \draw[#1] (#2) -- ++(.9,0) -- ++(-.9,-5) -- cycle; }
\newcommand\stdBlArrow{}
\def\stdBlArrow(#1)
  { \blArrow[line width=2pt, draw=red!70!black, fill=red!40, rounded corners=1pt](#1) }

\newcommand\redArrow{}
\def\redArrow(#1);
  { \draw[thick] ($(#1)+(-.3,.1)$) -- ($(#1)+(0,-.1)$) -- ($(#1)+(.3,.1)$);}
\newcommand\fSTS{}
\def\fSTS(#1);
  { \draw[fill=red!40, thick, draw=red!60!black] (#1) circle (.5);
    \draw[->] ($(#1)+(-.05,-.05)$) to ++(0,.4);
    \draw[->] ($(#1)+(-.05,-.05)$) to ++(.4,0);
    \draw[->] ($(#1)+(-.05,-.05)$) to ++(-.2,-.2);}
\newcommand\reduction{}
\def\reduction(#1);
  { \draw[fill=cyan!40, thick, draw=cyan!60!black] (#1) circle (.5);
    \draw[fill=black] ($(#1) + (-.2, .25)$) to ++(0.4,0) to ++(-.3,-.6) to cycle;}
\newcommand\customLegend{}
\def\customLegend(#1);
  {\fSTS(#1);
   \node[right of=#1] {fixed space-time structure};
   \draw[rounded corners=2pt] ($(#1)+(-.5,-.5)$) rectangle ($(#1)+(4, -4)$);}

\makeatletter
\newcommand*{\blitzset}{\pgfqkeys{/blitz}}\blitzset{
  height/.initial=4cm,
  width/.initial=2cm,
  breadth/.initial=.3cm,
  ratio/.initial=.4,
}

\pgfdeclareshape{blitz}
{
  \savedanchor\centerpoint{
    \pgf@x = .5\wd\pgfnodeparttextbox
    \pgf@y = .5\ht\pgfnodeparttextbox
  }
  \anchor{center}{\centerpoint}
  \anchor{north}{\centerpoint\pgfmathsetlength\pgf@ya{\pgfkeysvalueof{/blitz/ratio}*\pgfkeysvalueof{/blitz/height}}\advance \pgf@y by \pgf@ya}
  \anchor{south}{\centerpoint\pgfmathsetlength\pgf@ya{-(1-\pgfkeysvalueof{/blitz/ratio})*\pgfkeysvalueof{/blitz/height}}\advance \pgf@y by \pgf@ya}
  \anchor{east}{\centerpoint\pgfmathsetlength\pgf@ya{\pgfkeysvalueof{/blitz/breadth}}\advance \pgf@y by \pgf@ya\pgfmathsetmacro{\alpha}{atan(2*\pgfkeysvalueof{/blitz/ratio}*\pgfkeysvalueof{/blitz/height}/\pgfkeysvalueof{/blitz/width})}\pgfmathsetlength\pgf@xa{.5*\pgfkeysvalueof{/blitz/width}+\pgfkeysvalueof{/blitz/breadth}/tan(\alpha/2)}\advance\pgf@x by \pgf@xa}
  \anchor{west}{\centerpoint\pgfmathsetlength\pgf@ya{\pgfkeysvalueof{/blitz/breadth}}\advance \pgf@y by -\pgf@ya\pgfmathsetmacro{\alpha}{atan(2*\pgfkeysvalueof{/blitz/ratio}*\pgfkeysvalueof{/blitz/height}/\pgfkeysvalueof{/blitz/width})}\pgfmathsetlength\pgf@xa{.5*\pgfkeysvalueof{/blitz/width}+\pgfkeysvalueof{/blitz/breadth}/tan(\alpha/2)}\advance\pgf@x by -\pgf@xa}

  \backgroundpath{
    \centerpoint \pgf@xa=\pgf@x \pgf@ya=\pgf@y
    \pgfmathsetmacro{\alpha}{atan(2*\pgfkeysvalueof{/blitz/ratio}*\pgfkeysvalueof{/blitz/height}/\pgfkeysvalueof{/blitz/width})}
    \pgfmathsetlength\pgf@xb{.5*\pgfkeysvalueof{/blitz/width}-\pgfkeysvalueof{/blitz/breadth}/tan(\alpha/2)}
    \pgfmathsetlength\pgf@yb{\pgfkeysvalueof{/blitz/breadth}}
    \advance \pgf@xa by \pgf@xb
    \advance \pgf@ya by -\pgf@yb
    \pgfpathmoveto{\pgfpoint{\pgf@xa}{\pgf@ya}}
    \pgfmathsetlength\pgf@xb{\pgfkeysvalueof{/blitz/width}}
    \advance \pgf@xa by -\pgf@xb
    \pgfpathlineto{\pgfpoint{\pgf@xa}{\pgf@ya}}
    \pgfmathsetlength\pgf@yb{\pgfkeysvalueof{/blitz/breadth}+\pgfkeysvalueof{/blitz/ratio}*\pgfkeysvalueof{/blitz/height}}
    \pgfmathsetlength\pgf@xb{\pgf@yb*sin(90-\alpha)}
    \advance \pgf@xa by \pgf@xb
    \advance \pgf@ya by \pgf@yb
    \pgfpathlineto{\pgfpoint{\pgf@xa}{\pgf@ya}}
    \pgfmathsetlength\pgf@xb{2*\pgfkeysvalueof{/blitz/breadth}/cos(90-\alpha)}
    \advance \pgf@xa by \pgf@xb
    \pgfpathlineto{\pgfpoint{\pgf@xa}{\pgf@ya}}
    \pgfmathsetlength\pgf@yb{\pgfkeysvalueof{/blitz/ratio}*\pgfkeysvalueof{/blitz/height}-\pgfkeysvalueof{/blitz/breadth}}
    \pgfmathsetlength\pgf@xb{\pgf@yb*sin(90-\alpha)}
    \advance \pgf@xa by -\pgf@xb
    \advance \pgf@ya by -\pgf@yb
    \pgfpathlineto{\pgfpoint{\pgf@xa}{\pgf@ya}}
    \pgfmathsetlength\pgf@xb{\pgfkeysvalueof{/blitz/width}}
    \advance \pgf@xa by \pgf@xb
    \pgfpathlineto{\pgfpoint{\pgf@xa}{\pgf@ya}}
    \pgfmathsetlength\pgf@yb{(1-\pgfkeysvalueof{/blitz/ratio})*\pgfkeysvalueof{/blitz/height}+\pgfkeysvalueof{/blitz/breadth}}
    \pgfmathsetlength\pgf@xb{.5*\pgfkeysvalueof{/blitz/width}+\pgfkeysvalueof{/blitz/breadth}/tan(\alpha/2)}
    \advance \pgf@xa by -\pgf@xb
    \advance \pgf@ya by -\pgf@yb
    \pgfpathlineto{\pgfpoint{\pgf@xa}{\pgf@ya}}
    \pgfpathclose
  }
}

\newcommand*{\srefset}{\pgfqkeys{/sref}}\srefset{
  height/.initial=.5cm,
  width/.initial=.5cm,
}

\pgfdeclareshape{sref}
{
  \savedanchor\centerpoint{
    \pgf@x = .5\wd\pgfnodeparttextbox
    \pgf@y = .5\ht\pgfnodeparttextbox
  }
  \anchor{center}{\centerpoint}
  \anchor{north}{\centerpoint\pgfmathsetlength\pgf@ya{.5*\pgfkeysvalueof{/sref/height}}\advance \pgf@y by \pgf@ya}
  \anchor{south}{\centerpoint\pgfmathsetlength\pgf@ya{-.5*\pgfkeysvalueof{/sref/height}}\advance \pgf@y by \pgf@ya}
  \anchor{east}{\centerpoint\pgfmathsetlength\pgf@xa{.5*\pgfkeysvalueof{/sref/width}}\advance \pgf@x by \pgf@xa}
  \anchor{west}{\centerpoint\pgfmathsetlength\pgf@xa{-.5*\pgfkeysvalueof{/sref/width}}\advance \pgf@x by \pgf@xa}

  \backgroundpath{
    \centerpoint \pgf@xa=\pgf@x \pgf@ya=\pgf@y
    \pgfmathsetlength\pgf@xb{.5*\pgfkeysvalueof{/sref/width}}
    \pgfmathsetlength\pgf@yb{.5*\pgfkeysvalueof{/sref/height}}
    \pgfmathsetlength\pgf@yc{.75*\pgfkeysvalueof{/sref/height}}
    \advance \pgf@xa by \pgf@xb
    \pgfpathmoveto{\pgfpointpolar{20}{\pgf@xa}}
    \pgfpatharc{20}{170}{\pgf@xb}
    \pgfpathlineto{\pgfpointpolar{140}{\pgf@yc}}
    \pgfpathmoveto{\pgfpointpolar{200}{\pgf@xa}}
    \pgfpatharc{200}{350}{\pgf@xb}
    \pgfpathlineto{\pgfpointpolar{320}{\pgf@yc}}
  }
}
\makeatother

\makeatletter
\@ifclassloaded{beamer}
  {}
  {

    \theoremstyle{remark}
    
    \theoremstyle{definition}

  }
\makeatother

\newif\ifradical
\radicaltrue
\newif\ifbeta
\betatrue
\newif\ifuncertain
\uncertaintrue
\newif\ifcomments
\commentstrue

\newcounter{CtrSprachspiel}

\makeatletter
\def\p@paragraph{\thesection\,\thesubsection.}
\makeatother

\let\oldparagraph=\paragraph
\renewcommand\paragraph[1]{\oldparagraph{#1.}}

\makeatletter
\begin{document}

\title{The contextual state}
\title{Against the reified state}
\title{Against reification}
\title{Ceci n'est pas un syst\`{e}me isol\'{e}}
\title{Disturbance is not a bug, but a feature}
\title{Disturbance: It's a Feature, not a Bug}
\title{Contextuality: It's a Feature, not a Bug}

\author{Arne Hansen and Stefan Wolf}
\affiliation{Facolt\`a di Informatica, 
Universit\`a della Svizzera italiana, Via G. Buffi 13, 6900 Lugano, Switzerland}

\date{\today}

\begin{abstract}
\noindent
Results of measurements give legitimacy to a physical theory.
What if acquiring these results in the first place necessitates what the same theory considers to be an interaction?
In this note, we assume that theories account for interactions so that they are \emph{empirically traceable}, and that \emph{observations necessarily go with such an interaction} with the observed system.
We investigate consequences of this assumption: 
The unfolding language game, inspired by ``quantum logic,'' leads to a class of contextual and probabilistic theories.
Contextuality becomes a means to render interactions, thus also measurements, empirically tangible. 
The measurement ``problem'' arises in all such theories, not only quantum mechanics:
It is a consequence of the need for empirical evidence of interactions.
And a consequence of the converse need for an interaction when obtaining that empirical evidence. 

 \end{abstract}

\maketitle

\section{Introduction}
\label{sec:introduction}
\noindent
The infamous Wigner's-friend experiment~\cite{wigner1963problem, deutsch1985quantum, Wigner1961} serves to illustrate the measurement problem:
If we imagine Wigner performing a measurement on his friend who measured another system, there are different---in fact, incommensurable---uses of the term ``measurement:''
\begin{enumerate}[label={(M\arabic*)}]
  \item\label{M1} If the friend's ``measurement'' of a state in an equal superposition with respect to his measurement basis is regarded as an interaction between two systems---modelled by a physical evolution---, then it corresponds to a \emph{unitary} on the joint system, yielding an entangled joint state;
  \item\label{M2} if the ``measurement'' leads to an account of experience that serves as a normative judgement about the validity of a theory, we expect \emph{exclusively} one of several possible outcomes.
\end{enumerate}
Statements~\ref{M1} and~\ref{M2} are in conflict in two respects:
On the one hand, there is a fundamental separation due to the nature of language: 
A meaningful account of experience cannot be exhaustively captured by an interaction as expressed in the language of one theory~\cite{HW18}.
On the other hand, the linearity of quantum mechanics cannot be reconciled with value-definiteness---i.e., the outcome being exclusively one of several possibilities.

The latter incommensurability is not so much a peculiarity---or defect---of quantum mechanics.
Instead, we argue that it appears in theories that 
\begin{enumerate*}[label={(\alph*)}]
  \item account for interactions so that they are empirically significant, and 
  \item require that an observation necessarily goes with \emph{such} an interaction.
\end{enumerate*}
An observation is then itself empirically traceable.
The two requirements above are combined in the \emph{interaction assumption}: 
\begin{enumerate}[label={(IntA)}]
  \item\label{IntAssum} Interactions are empirically traceable. An observation necessitates such an interaction.
\end{enumerate}
The incommensurability in the measurement problem can, therefore, be regarded as a \emph{consequence of the interaction assumption}.

In order to determine the accounts of experience that are compatible quantum mechanics---or, conversely, the accounts that falsify quantum mechanics---, we have to connect statements~\ref{M1} and~\ref{M2}.
If they are incommensurable, then a bridge is needed: the Born rule.
The necessity of a ``Born rule'' is tightly connected to the measurement problem and can as well be traced back to the interaction assumption~\ref{IntAssum}.
\emph{We argue that both appear in a wider class of physical theories that satisfy the interaction assumption, not just quantum mechanics.}

\iffalse
\begin{figure}
  \centering
  \begin{tikzpicture}[thick,scale=1.1, every node/.style={transform shape}]
    \def\d{-1.2}
    \node[colorbox=red] (born) at (0, .8*\d) {Born rule};
\node[font=\footnotesize, inner sep=0pt, anchor=north east] (nec) at ($(born.south east)+(0,-.05)$) {necessity};
\node[colorbox=blue] (measure) at (0, -.8*\d) {measurement problem};
\node[font=\footnotesize, inner sep=0pt, anchor=south east] (incom) at ($(measure.north east)+(0,.05)$) {incommensurability};
\draw[conn=4pt, ->] (born) to[bend left] (measure);
\draw[conn=4pt, <-] (born) to[bend right] (measure);
\node[colorboxS=violet, rotate=10] (interaction) at (0,0) {interaction assumption};
   \end{tikzpicture}
  \caption{There is a connection between the necessity of the Born rule and the incommensurability exposed in the measurement problem. 
  They both arise from the incommensurability of statements~\ref{M1} and~\ref{M2}.
  The two can be regarded as arising from the interaction assumption.}
  \label{fig:born_interaction_measureProb}
\end{figure}
\fi

\iffalse
Despite being inspired by quantum mechanics, our discussion revolves around physical theories in general.
We use terms that have a similar meaning in the context of quantum mechanics. 
Unless explicitly stated, the meaning of terms in our, more general, context does, however, \emph{not derive from} quantum mechanics. 
\fi

\begin{figure}
  \centering
  \begin{tikzpicture}
    \def\d{-1.2}
    \def\h{3.2}
    \node[colorbox=orange] (theory) at (0, \d) {theories};
\draw[conn=2pt, rounded corners] (theory.north) to ++(0,.2) to ++(-.4*\h, -.2*\d) to ++(0,.2) 
  coordinate (c11);
\node[colorbox=violet, anchor=south] (intAssum) at ($(c11)+(-.1*\h, 0)$) {interaction assumption};
\draw[conn=2pt, rounded corners] (theory.north) to ++(0,.2) to ++(.4*\h, -.2*\d) to ++(0,.2) coordinate (c12);
\node[colorbox=violet, anchor=south] (verInf) at ($(c12)+(.1*\h, 0)$) {verifiability};
\coordinate (c21) at ($(intAssum.north)+(.2*\h, 0)$);
\coordinate (c22) at ($(verInf.north)+(-.2*\h, 0)$);
\node[anchor=south west, inner sep=0, font=\footnotesize] at ($(intAssum.north west)+(0,.1)$) {Sec.~\ref{sec:meaning_interaction_sys}};
\node[anchor=south east, inner sep=0, font=\footnotesize] at ($(verInf.north east)+(0,.1)$) {Sec.~\ref{sec:veridical_information}};
\node[inner sep=0, font=\footnotesize] (sec) at (intAssum |- theory) {Sec.~\ref{sec:complementarity},~\ref{sec:interactions_within_isolated_systems}};
\draw[gconn=4pt, densely dotted] (sec) to (intAssum);
\draw[gconn=4pt, densely dotted] (sec) to (theory);
\draw[conn=2pt, densely dotted] (theory.south) to node[midway, right, font=\footnotesize] {App.~\ref{sec:state_system}} ++(0,.4*\d) node[colorbox=yellow,anchor=north, solid] {$\mathcal{Q}$};
   \end{tikzpicture}
  \caption{We investigate how the interaction assumption and a concept of verifiability can be combined.}
  \label{fig:overview}
\end{figure}
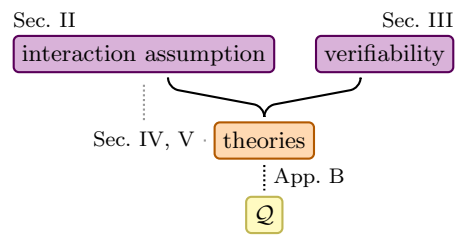

The structure of the article is illustrated in Figure~\ref{fig:overview}:
Our understanding of an ``interaction,'' relies on the notion of a ``system'' which we discuss in Section~\ref{sec:meaning_interaction_sys}.
In Section~\ref{sec:veridical_information}, we develop the notion of \emph{verifiability}, a concept constitutive to natural sciences. 
We then investigate in Sections~\ref{sec:complementarity} and~\ref{sec:interactions_within_isolated_systems} how the notion of verifiable information can be reconciled with empirically tangible interactions as required by~\ref{IntAssum}.
In Section~\ref{sec:lattice_questions}, we join the equivalence relation arising from the demand for verifiability with the lattice structure inspired by quantum logic~(see Appendix~\ref{sec:qmlogics}).
In Appendix~\ref{sub:resolution_restriction}, we show how the resolution restriction emerges in theories with an interaction assumption, and in Appendix~\ref{sec:state_system}, we discuss repercussions of the above for quantum mechanics. 
 \section{Systems and interactions}
\label{sec:meaning_interaction_sys}
\begin{figure}[h]
  \centering
  \begin{tikzpicture}
    \def\d{-.4}
\node[colorbox=cyan] (isolSystem) at (0, \d) {system};
\node[colorbox=orange] (equivQuest) at (0,-\d) {interaction};
\draw[conn=4pt, ->] (isolSystem.east) .. controls ($(isolSystem.east)+(.8,0)$) and ($(equivQuest.east)+(.8,0)$) .. (equivQuest.east);
\draw[conn=4pt, <-] (isolSystem.west) .. controls ($(isolSystem.west)+(-.8,0)$) and ($(equivQuest.west)+(-.8,0)$) .. (equivQuest.west);
   \end{tikzpicture}
  \caption{The notion of an interaction and the notion of a (isolated) system are closely related.}
  \label{fig:int_sys}
\end{figure}
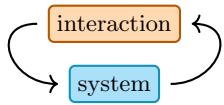

\noindent
In the following, we contextualize the interaction assumption~\ref{IntAssum}: Interactions are understood as a particular way in which two \emph{systems} relate to one another~(see Figure~\ref{fig:int_sys}).
In Sections~\ref{sub:separability} and~\ref{sub:the_notion_of_a_system}, we thus examine the notion of a ``system:''
On the one hand, the scientific quest for empirical confirmation and reproducibility requires a context in which statements can only depend on that context.
On the other hand, as discussed in Section~\ref{sub:no_action_at_a_distance}, if an observation is not a ``spooky action at a distance,'' the observer and the observed entity cannot be regarded as independent.

\subsection{Separability}\label{sub:separability}
\noindent
The natural sciences rely essentially on empirically confirming statements---as we discuss in greater detail in Section~\ref{sec:veridical_information}.
We, therefore, require that there is a context in which there are statements with a clearly confined dependence.
Einstein's separability assumption---effectively a \emph{non-signalling assumption}---ensures that we can meaningfully say something about an ``entity'' or ``system,'' \emph{independently} of its environment, that is, independently of parts external to the system:\footnote{Quantum mechanics accounts for this by the partial trace being sufficient to derive all measurement results about the subsystem (see, e.g.,~\cite{QIT_script}).}
 \blockcquote[\S5 (translated quote from~\cite{EinsteinBorn})]{sep-einstein-philo}[.]{
   that which we conceive as existing (`actual') should somehow be localized in time and space. 
  That is, the real in one part of space,~$A$, should (in theory) somehow `exist' independently of that which is thought of as real in another part of space,~$B$. 
  If a physical system stretches over the parts of space~$A$ and~$B$, then what is present in~$B$ should somehow have an existence independent of what is present in~$A$. 
  What is actually present in~$B$ should thus not depend upon the type of measurement carried out in the part of space,~$A$; it should also be independent of whether or not, after all, a measurement is made in~$A$}

The above contradicts the experimental findings on non-locality~\cite{aspect1982experimental, ma2012experimental} if separability is turned into a sufficient condition in the following sense:
Measurements on parts, together with previously shared information, reveal the results of any measurement that can possibly be performed on the combined system. We must, therefore, allow for measurements on the combined system that cannot be characterized by measurements on the parts together with previously shared information.

\subsection{The notion of a ``system''}\label{sub:the_notion_of_a_system}
\noindent
Einstein combines the assumption of systems being separately describable with the observation of the ``worldly origin'' of our terms:\footnote{\blockcquote[p.~102, own translation]{Einstein1916}[.]{Terms that have proven useful for the ordering of things attain easily such an authority over us so that we forget their worldly origin and we accept them as unalterable facts.
They are, then, put down as `thinking-necessities,' `a priori given,' etc.
The path of scientific progress is often made impassable for a long time by such misconceptions}
 } 
\blockcquote[\S5 (translated quote from~\cite{EinsteinBorn})]{sep-einstein-philo}[.]{However, if one renounces the assumption that what is present in different parts of space has an independent, real existence, then I do not at all see what physics is supposed to describe. 
  For what is thought to be a `system' is, after all, just conventional, and I do not see how one is supposed to divide up the world objectively so that one can make statements about the parts}
If one abstains from supposing privileged linguistic means that ``define'' what are ``the systems,'' then we are lead to allow for an unrestricted choice of what to consider as a system. 
Thus, in light of the considerations in Section~\ref{sub:separability}, we require that whatever we \emph{call} a system has an independent description:
\emph{Our choice of a ``system'' should not affect the ability to formulate independent statements about that system.}

Whereas a theory with this flexibility as to what counts as a ``system'' allows for ample applications, it deprives the term of its sortal character and thus its ability to establish \emph{identity}:\footnote{\blockcquote[\S2.2]{glock2003}[.]{A \emph{criterion} is a way of recognizing something, or a feature by which we can recognize something. A criterion of \emph{identity} is something by which we can recognize the correctness of a statement of identity. 
Questions of identity make sense only once we have specified what kind of thing is at issue.
One cannot simply ask `Is this the same as that?', but only `Is this the same~$S$ as that', where `$S$' is a sortal term like `table' or `planet'. 
Accordingly, a criterion of identity is a feature which determines whether or not an object falling under `$S$' that we encounter in one context or refer to in one way is the same as an object falling under `$S$' that we encounter in another context or refer to in another way}
 }
\emph{If \textbf{anything} can be regarded as a system, \textbf{nothing} is essentially a system.}
The mere character trait of ``constituting a system'' does not provide the means to identify the system.
To meaningfully say something about a system, however, requires us to identify and refer to that system.\footnote{\blockcquote[\S1]{strawson2000}[.]{Another possible interpretation of the slogan [`no entity without identity'] [\ldots] might run something like this: `There is nothing you can sensibly talk about without knowing, at least in principle, how it might be identified.' I have nothing to say against \emph{this} admirable maxim}
 }
So, a theory that states \emph{what} can be said about a system, and that comes with the flexibility regarding the choice of the system, necessarily relies on other linguistic means to establish the reference to that system.\footnote{In~\cite{Mittelstaedt2005}, Mittelstaedt proposes a ``quantum ontology.'' 
The author rightly points out the inability to re\"{i}dentify a ``quantum object.''
Thus, the ``quantum ontology'' cannot suffice to meaningful refer to either a system or a ``quantum object.''
Even if quantum mechanics is \textcquote[p.9]{Mittelstaedt2005}[]{nearer to the `truth' than classical mechanics}, it cannot be the ``whole truth.''}
Furthermore, if a theory does not allow to identify its systems, it does also not allow to identify any potential thing-in-itself associated with systems~(see also Appendix~\ref{sec:state_system}).

\subsection{No observation at a distance}\label{sub:no_action_at_a_distance}
\noindent
Despite the possibility of statements independent of the environment of a system, we have to allow for changes of a system that are \emph{the effect of an external cause}.
Regarding an observer as a system, we have to account for the external cause of his sensory perceptions. This does \emph{neither} imply that this description of the observer is exhaustive, \emph{nor} imply the ability to conclude what is the observer's account of experience~\cite{HW18}. 
It merely requires an interaction as the necessary requirement for the external cause of his sensory perception.
We, thus, restrict separability in the sense that \emph{a system is either independently describable---it is \textbf{isolated}---, or it \textbf{interacts} with its environment---i.e., with other systems external to it.}
There is, therefore, a dichotomy between systems that are independently describable \emph{versus} systems that interact with their environment---with such an interaction, seemingly contradictory, being at the core of ``saying something about a system.''

The interaction assumption~\ref{IntAssum} demands that these interactions themselves have empirically detectable effects.
\emph{An interaction is, therefore, not an abstract term beyond our experience but it becomes itself meaningful.}
If an observation is necessarily accompanied by an interaction, it must be empirically detectable as well.
\emph{There is no ``observation at a distance,'' just as there is no action at a distance.}

The importance of an interaction has been emphasized before. 
Bohr, in his reply~\cite{Bohr1935} to the EPR paper~\cite{EPR}, emphasizes the role of interactions to refute the notion of a \textcquote[]{EPR}{[prediction] without without in any way disturbing a system}---the idea of innocently reading off measurement results:\footnote{Dewey observes the one-sided effect of an ``interaction:'' \blockcquote[{\S}1, p.26]{DeweyQFC}[.]{The theory of knowing is modelled after what was supposed to take place in the act of vision.
The object refracts light to the eye and is seen; it makes a difference to the eye and to the person having an optical apparatus, but none to the thing seen. The real object is the object so fixed in its regal aloofness that it is a king to any beholding mind that may gaze upon it.
A spectator theory of knowledge is the inevitable outcome}
 }
\blockcquote[]{Bohr1935}[.]{Indeed the \emph{finite interaction between object and measuring agencies} conditioned by the very existence of the quantum of action entails---because of the impossibility of controlling the reaction of the object on the measuring instruments if these are to serve their purpose---the necessity of a final renunciation of the classical ideal of causality and a radical revision of our attitude towards the problem of physical reality}
 In a similar vein, Popper in~\cite{PopperIndet1} exposes an essential shortcoming of classical mechanics: If classical mechanics did account for the interactions that finally led to our experiences---if it did satisfy the interaction assumption~\ref{IntAssum}---, then it would be indeterministic.
Instead, classical mechanics relies on \emph{other} theories to account for causal connections in an observation: 
The electromagnetic interactions allowing to measure position and momentum have negligible disturbing effects.\footnote{In~\cite{Aerts83}, Aerts distinguishes between ``theories of system'' and ``theories of measurements.''
The former ``predict'' results of measurements. 
The latter account for disturbances by the measurement.
Classical and quantum mechanics both belong to the former. 
Complementarity is then, according to Aerts, not a disturbance by a measurement, but the \emph{inability to predict with certainty}.
We neither follow the distinction regarding theories nor the author's statement that 
\blockcquote[p.2442]{Aerts83}[.]{The aim of a physical theory of the physical system is to `predict' the result of a certain test, and this prediction is done before the test is carried out and no matter whether the test will be carried out}
 This characterization of the aim of physics requires an a priori meaning of a ``certain test'' independent from running that test, contrary to our belief that meaning is established within the context of a language game.}

 \section{Verifiability}\label{sec:veridical_information}
\iffalse
\noindent
In Section~\ref{sub:equivalent_questions}, we discuss the need for an equivalence of questions in order to allow for a verifiability.
In Section~\ref{sub:formal_description} we introduce a formal representation of questions as an orthocomplemented lattice.
The lattice is then, in Sections~\ref{sub:formal_description} and~\ref{sub:the_system}, endowed with the equivalence relations as depicted in Figure~\ref{fig:equiv_quest}.
\fi

\subsection{Equivalent questions}\label{sub:equivalent_questions}
\noindent
If physics, in the spirit of Einstein's quote in Section~\ref{sub:the_notion_of_a_system} above, strives to make ``statements about parts of the world,'' then we require that statements can repeatedly be empirically confirmed---they are \emph{verifiable}.
Popper states in~\cite{Popper1934} that the ``scientifically meaningful physical effect'' is characterized by being \emph{reproducible}---\emph{regularly} and by \emph{anybody who builds the experiment according to the instructions}.\footnote{\blockcquote[{\S}I.8]{Popper1934}[.]{Der wissenschaftlich belangvolle physikalische Effekt kann ja geradezu dadurch definiert werden, da{\ss} er sich regelm\"{a}{\ss}ig und von jedem reproduzieren l\"{a}{\ss}t, der die Versuchsanordnung nach Vorschrift aufbaut}
 }
This supposes that there exists the possibility to \emph{refer to} or \emph{mean}  the ``same experiment'' for different points in space and time (``regularly''), and by different observers (``anybody''). We, therefore, assume:\footnote{We do not specify how this equivalence is established~(see also last paragraph in Section~\ref{sub:isolated_system}).}
\begin{enumerate}[label={(EQ)}]
  \item\label{EquivQuest} There is an \emph{equivalence} between questions.
    If and only if an observation yields an \emph{equal answer} to an \emph{equivalent question}, then the second answer \emph{confirms} the first.\end{enumerate}
The assumption does not imply that there is an a priori fixed meaning of ``experiment built according to \emph{the} instructions'' captured in a particular privileged language.
The assumption~\ref{EquivQuest} of the possibility to ask semantically equivalent questions can be read as: 
There is the possibility to \emph{establish} identical meaning and thus equivalent questions for both different points in space and time, as well as different observers.

While we assume that there are equivalent questions, we do not assume that equivalent questions always have equal answers.
Rather, we use in Section~\ref{sub:isolated_system} that answers can \emph{depend on the context}, i.e., they can be \emph{contextual}.

To capture the above formally, we assume that questions in a given experimental context are represented by elements of a set~$\mathcal{Q}$.
Two questions $Q_1, Q_2\in\mathcal{Q}$ are equivalent in the above sense~(see Section~\ref{sub:equivalent_questions}) if and only if~$Q_1$ and~$Q_2$ are $\sim$-equivalent, i.e.,~$Q_1\sim Q_2$.
If we ``ask the same question twice,'' we are referring to two \emph{non-equal} $\sim$-equivalent questions
\begin{equation*}
  Q_1,Q_2\in\mathcal{Q}: Q_1 \sim Q_2 \ \land \  Q_1\neq Q_2 \,.
\end{equation*}
Thus, the elements in~$\mathcal{Q}$ have usually a time-stamp or a similar parameter that allows for a distinction between, e.g., the question ``what is the momentum at time~$t_1$'' and the question ``what is the momentum at time~$t_2$'' where~$t_1\neq t_2$.

We assume another set~$\mathcal{A}$ of answer to questions in~$\mathcal{Q}$. 
We write~$(Q,A)$ to associate the answer~$A$ with a question~$Q$.
For now, we assume the elements in~$\mathcal{Q}$ to correspond to binary questions, i.e., questions with two possible answers,~$\Bt$ and~$\Bf$. 
Thus, the set of answers is~$\mathcal{A}=\{\Bt, \Bf\}$.

Theories associate answers to questions.
For now, we take a theory to be deterministic in the sense that it associates each question~$Q\in\mathcal{Q}$ with an answer~$A\in\{\Bt,\Bf\}$. 
The association of an answer to a question may be contextual in the sense that the theory associates answers to any finite subset of questions $\overline{Q}\subset\mathcal{Q}$ \emph{dependent on that subset}.
Conversely, a theory is called \emph{non-contextual} if it associates answers to questions independently of other questions~(see Section~\ref{sub:born_rule_in_a_contextual_theory}).

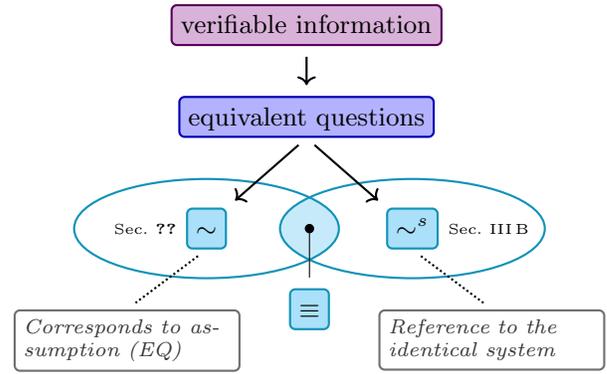
\begin{figure}
  \centering
  \begin{tikzpicture}[thick,scale=1.1, every node/.style={transform shape}]
    \def\d{-1.2}
    \def\h{3.2}
    \node[colorbox=violet] (verInf) at (0,0) {verifiable information};
\draw[conn=4pt, ->] (verInf.south) to ++(0,.5*\d) node[anchor=north, colorbox=blue] (eq) {equivalent questions};
\draw[conn=4pt, ->] (eq.south) to ++(-.3*\h,.7*\d) node[anchor=north east, colorbox=cyan] (equiv) {$\sim$};
\node[font=\tiny, anchor=east,inner sep=0] at ($(equiv.west)+(-.10,0)$) {Sec.~\ref{sub:formal_description}};
\draw[conn=4pt, ->] (eq.south) to ++(.3*\h,.7*\d) node[anchor=north west, colorbox=cyan] (equivS) {$\sim^s$};
\node[font=\tiny, anchor=west,inner sep=0] at ($(equivS.east)+(.10,0)$) {Sec.~\ref{sub:the_system}};
\begin{pgfonlayer}{bg}
  \begin{scope}
    \clip  (equivS) ellipse (1.6cm and .6cm);
    \fill[cyan!20] (equiv) ellipse (1.6cm and .6cm);
  \end{scope}
  \draw[cyan!70!black,thick] (equivS) ellipse (1.6cm and .6cm);
  \draw[cyan!70!black,thick] (equiv) ellipse (1.6cm and .6cm);
\end{pgfonlayer}
\draw[fill=black] ($.5*(equiv)+.5*(equivS)$) circle (.05);
\draw[shorten >=4pt] ($.5*(equiv)+.5*(equivS)$) to ++(0,.6*\d) node[anchor=north, colorbox=cyan] {$\equiv$};
\draw[densely dotted, conn=4pt] (equiv.south) to ++(-.3*\h,.6*\d) node[anchor=north, comment=2.5cm,solid] {Corresponds to assumption \ref{EquivQuest}};
\draw[densely dotted, conn=4pt] (equivS.south) to ++(.3*\h,.6*\d) node[anchor=north, comment=2.5cm,solid] {Reference to the identical system};
   \end{tikzpicture}
  \caption{Requiring verifiability leads to an equivalence of questions~($\sim$). 
  Further, we introduce an equivalence relation for questions about the identical system~($\sim^s$).
  The intersection of the two, $\sim$-equivalent questions about the identical system, is denoted by~$\equiv$.}
  \label{fig:equiv_quest}
\end{figure}

\subsection{Identical systems}\label{sub:the_system}
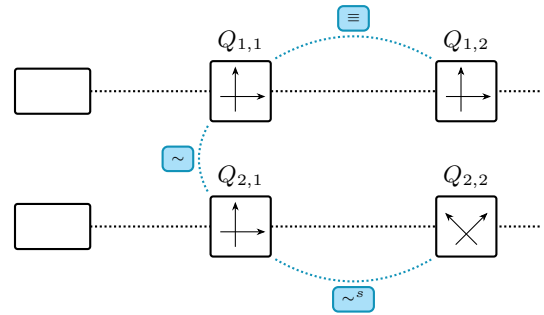
\begin{figure}[h]
  \centering
  \begin{tikzpicture}
    \def\w{.8}
\begin{scope}[shift={(0,1.8)}]
  \coordinate (m1) at (2,0);
  \coordinate (m2) at (5,0);
  \draw[thick, densely dotted] (0,0) to (6,0); 
  \draw[rounded corners=1pt, thick] (0,-.3) rectangle ++(-1,.6); 
  \draw[rounded corners=1pt, thick, fill=white] ($(m1)+(-.4,-.4)$) coordinate(m11) rectangle ++(\w,\w); 
  \draw[-{Stealth [scale=.7]}] ($(m1)+(225:.1)+(270:.2)$) to ++(.0,.6); 
  \draw[-{Stealth [scale=.7]}] ($(m1)+(225:.1)+(180:.2)$) to ++(.6,0); 
  \draw[rounded corners=1pt, thick, fill=white] ($(m2)+(-.4,-.4)$) coordinate(m12) rectangle ++(.8,.8); 
  \draw[-{Stealth [scale=.7]}] ($(m2)+(225:.1)+(270:.2)$) to ++(.0,.6); 
  \draw[-{Stealth [scale=.7]}] ($(m2)+(225:.1)+(180:.2)$) to ++(.6,0);
  \node[anchor=south] (nm1) at ($(m1)+(.0,.4)$) {$Q_{1,1}$};
  \node[anchor=south] (nm2) at ($(m2)+(.0,.4)$) {$Q_{1,2}$};
\end{scope}
\begin{scope}[shift={(0,0)}]
  \coordinate (m1) at (2,0);
  \coordinate (m2) at (5,0);
  \draw[thick, densely dotted] (0,0) to (6,0); 
  \draw[rounded corners=1pt, thick] (0,-.3) rectangle ++(-1,.6); 
  \draw[rounded corners=1pt, thick, fill=white] ($(m1)+(-.4,-.4)$) coordinate(m21) rectangle ++(.8,.8); 
  \draw[-{Stealth [scale=.7]}] ($(m1)+(225:.1)+(270:.2)$) to ++(.0,.6); 
  \draw[-{Stealth [scale=.7]}] ($(m1)+(225:.1)+(180:.2)$) to ++(.6,0);
  \draw[rounded corners=1pt, thick, fill=white] ($(m2)+(-.4,-.4)$) coordinate(m22) rectangle ++(.8,.8); 
  \draw[-{Stealth [scale=.7]}] ($(m2)+(270:.1)+(315:.2)$) to ++(135:.6); 
  \draw[-{Stealth [scale=.7]}] ($(m2)+(270:.1)+(225:.2)$) to ++(45:.6);
  \node[anchor=south] (nm1) at ($(m1)+(.0,.4)$) {$Q_{2,1}$};
  \node[anchor=south] (nm2) at ($(m2)+(.0,.4)$) {$Q_{2,2}$};
\end{scope}
\draw[cyan!70!black,thick,densely dotted, conn=2pt] (m11) to[bend right] node[midway,left=2pt,solid,black,colorboxXS=cyan] {$\sim$} ($(m21)+(0,\w)$);
\draw[cyan!70!black,thick,densely dotted, conn=2pt] ($(m11)+(\w,\w)$) to[bend left] node[midway,above=2pt,solid,black,colorboxXS=cyan] {$\equiv$} ($(m12)+(0,\w)$);
\draw[cyan!70!black,thick,densely dotted, conn=2pt] ($(m21)+(\w,0)$) to[bend right] node[midway,below=2pt,solid,black,colorboxXS=cyan] {$\sim^s$} ($(m22)+(0,0)$);
   \end{tikzpicture}
  \caption{Imagine, for instance, different runs of a Stern-Gerlach experiment. 
  Within each run, there are two measurement. 
  The bases in each of these measurements can be chosen. 
  Thus, the set of questions is~$\mathcal{Q}=\{(A,i,j)\vert A\in\End(\mathbb{C}^2),A^{\dagger}=A,i\in\mathbb{N}^+, j\in\{0,1\}\}$. 
  Two measurements are $\sim$-equivalent if the measurement bases coincide. 
  Two measurements are $\sim^s$-equivalent if they are in the same run, i.e., if the corresponding questions have the same first index.
  If in the same run two questions are $\sim$-equivalent, then they are $\equiv$-equivalent.}
  \label{fig:identical_systems}
\end{figure}

\noindent
In two subsequent measurements \emph{in one run} of a Stern-Gerlach experiment, we assume that we refer to two measurements of the same silver atom---i.e., of the \emph{identical entity} as depicted in Figure~\ref{fig:identical_systems}. 
Observations then yield attributes of one single entity.
\emph{Thus, two questions may be equivalent insofar as they refer to the identical entity.}
This yields a second equivalence relation,~$\sim^s$, on~$\mathcal{Q}$.
While questions equivalent with respect to $\sim$ may refer to the \emph{same kind of entity} in, e.g., different runs of the same experimental setup, questions that are equivalent with respect to~$\sim^s$ refer to the same entity within the one particular run.

The equivalence relation of $\sim$-equivalent questions referring to the identical system, i.e., the intersection of $\sim$ and $\sim^s$, is then denoted as~$\equiv$.
In this sense, performing two measurements in the same basis does not constitute asking the \emph{same} question but asking \emph{equivalent} questions with respect to the relation~$\equiv$.

In summary, we have introduced two notions of equivalence of questions that reflect essentially asking the same question about the same (type of) entity,~$\sim$, and about the same (identical) entity,~$\equiv$.
The latter corresponds to asking $\sim$-equivalent questions referring to the identical system, i.e., that are $\sim^s$-equivalent.
 \section{Interaction assumption for~verifiable~information}\label{sec:complementarity}
\noindent
We combine the notion of verifiable information established in Section~\ref{sec:veridical_information} with the interaction assumption~\ref{IntAssum} motivated in Section~\ref{sec:meaning_interaction_sys}.
First, in Section~\ref{sub:isolated_system}, we contrast \emph{isolated systems} against systems that \emph{interact with their environment}.
This leads us to turn to \emph{probabilistic and contextual theories}.
As discussed in Section~\ref{sub:born_rule_in_a_contextual_theory}, a \emph{Born rule} then becomes an essential part of the theory.
In Section~\ref{sub:general_bb84}, we examine connections to the BB84 key-agreement protocol.
Subsequently, in Section~\ref{sec:interactions_within_isolated_systems}, we consider interactions between different parts of an isolated system, and how they relate to the measurement problem.

\subsection{Interactions and isolated systems}\label{sub:isolated_system}
\noindent

\begin{figure}[h]
  \centering
  \begin{tikzpicture}
    \def\h{1.5}
\def\d{-1}
\draw[thick, densely dotted] (-.5*\h, 0) coordinate (e1) to (2.5*\h,0) coordinate (e2);
\draw[<-, shorten <=2pt] (e1) to[bend left=25] ++(-.6*\h,-.45*\d) node[anchor=south] {``the system''};
\node[colorbox=blue] (q1) at (0,0) {$Q_1$};
\node[colorbox=blue] (q2) at (2*\h,0) {$Q_2$};
\draw[conn=2pt, thick, ->] (q1) to ++(0,.7*\d) node[anchor=north] {$A_1$};
\draw[conn=2pt, thick, ->] (q2) to ++(0,.7*\d) node[anchor=north] {$A_2=A_1$};
\draw[conn=2pt, thin] (q1.north east) to[bend left] node[colorboxXS=cyan,above=2pt] {$\equiv$} (q2.north west);
   \end{tikzpicture}
  \caption{In an isolated system, $\equiv$-equivalent questions about the identical system, $Q_1\equiv Q_2$~(i.e.,~$Q_1\sim Q_2 \land Q_1\sim^s Q_2$), yield the same answer~$A_1=A_2$.
    The reference to the same system is indicated by the dotted line.}
  \label{fig:isolated_det_syst}
\end{figure}
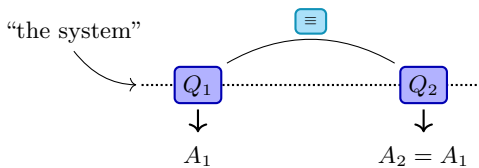

\paragraph{Isolated systems}\label{par:isolated_systems}
We first characterize what it means for a system \emph{not} to interact with its environment: 
\emph{A system~$S$ is \textbf{isolated} if and only if equivalent questions referring to that system yield same answers.}
We follow the intuition that ``being isolated'' is a property of a particular system~$S$.
Thus, the notion of being isolated relies merely on questions in one particular equivalence class~$\tilde{Q}^s$---of~$\sim^s$-equivalent questions referring to~$S$:
\begin{IEEEeqnarray}{RL}\label{eqn:isol_det}
  &S \ \text{is \emph{isolated} if and only if} \\
  &\forall Q \in \tilde{Q}^s \ \text{with} \ (Q,A): \forall Q' \equiv Q: (Q', A) \, . \nonumber
\end{IEEEeqnarray}
For isolated systems, $\equiv$-equivalent questions ($\sim$- and $\sim^s$-equivalent, see Section~\ref{sub:the_system}) obtain the same answers~(see Figure~\ref{fig:isolated_det_syst}).

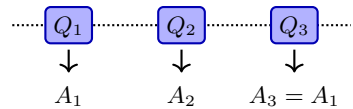
\begin{figure}[h]
  \centering
  \begin{tikzpicture}
    \def\h{1.5}
\def\d{-1}
\draw[thick, densely dotted] (-.5*\h, 0) to (2.5*\h,0) coordinate (e1);
\node[colorbox=blue] (q1) at (0,0) {$Q_1$};
\node[colorbox=blue] (q2) at (\h,0) {$Q_2$};
\node[colorbox=blue] (q3) at (2*\h,0) {$Q_3$};
\draw[conn=2pt, thick, ->] (q1) to ++(0,.7*\d) node[anchor=north] {$A_1$};
\draw[conn=2pt, thick, ->] (q2) to ++(0,.7*\d) node[anchor=north] {$A_2$};
\draw[conn=2pt, thick, ->] (q3) to ++(0,.7*\d) node[anchor=north] {$A_3=A_1$};

   \end{tikzpicture}
  \caption{By transitivity of the equivalence relation~$\sim$, asking an equivalent question~$Q_2 \equiv Q_1$ does not disturb the system: The answer to another equivalent question~$Q_3\equiv Q_1$---short for $Q_2\sim Q_1\land Q_1\sim^s Q_2$---is still the same,~$A_3=A_1$.}
  \label{fig:equiv_questions}
\end{figure}

\paragraph{Transitivity}\label{par:transitivity}
We take the relations~$\sim$ and~$\sim^s$ to be proper equivalence relations, in particular \emph{transitive}.
To ensure consistence with our notion of \emph{isolated systems}, an interaction that reproduces valid information---that yields an answer to a previously asked equivalent question---leaves an isolated system undisturbed.
An isolated system may interact with its environment to inquire about equivalent questions without being disturbed as depicted in Figure~\ref{fig:equiv_questions}.
In other words: \emph{By transitivity we cannot distinguish whether a system is not interact with at all or whether it is interacted with to inquire about an equivalent question.}

\paragraph{Empirically tangible interactions}\label{par:empirically_tangible_interactions}
The interaction assumption demands that we are able to distinguish between a system that is isolated and a system that has been interacted with to inquire non-implied information.\footnote{The action of an eavesdropper can then be detected as we will discuss in Section~\ref{sub:general_bb84}.}
We extend the interaction assumption~\ref{IntAssum} slightly: Not only do we assume that there is an interaction corresponding to every inquiry about a system. 
For every interaction, there is also a corresponding question an observer equivalently could have inquired about.\footnote{In fact, requiring that there should not be any qualitative difference between interactions can be read this way.}
We, therefore, treat all interactions of a system with its environment as if an observer is inquiring about a question in~$\mathcal{Q}$.

\begin{figure}[h]
  \centering
  \begin{tikzpicture}
    \def\h{1.5}
\def\d{-1}
\draw[thick, densely dotted] (-.5*\h, 0) to (2.5*\h,0) coordinate (e1);
\node[colorbox=blue] (q1) at (0,0) {$Q_1$};
\node[colorbox=orange] (q2) at (\h,0) {$Q_2$};
\node[colorbox=blue] (q3) at (2*\h,0) {$Q_3$};
\draw[conn=2pt, thick, ->] (q1) to ++(0,.7*\d) node[anchor=north] {$A_1$};
\draw[conn=2pt, thick, ->] (q2) to ++(0,.7*\d) node[anchor=north] {$A_2$};
\draw[conn=2pt, thick, ->] (q3) to ++(0,.7*\d) node[anchor=north] {$A_3\neq A_1$};
\draw[conn=2pt, thin] (q1.north east) to[bend left] node[colorboxXS=cyan,above=2pt] {$\equiv$} (q3.north west);
   \end{tikzpicture}
  \caption{Asking a non-equivalent question about a system disturbs it in the sense that equivalent questions about it, $Q_1\equiv Q_3$, yield different answers.}
  \label{fig:disturbance_detTheory}
\end{figure}
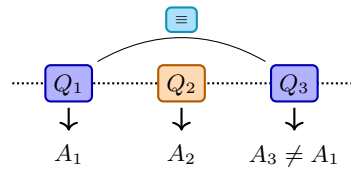

\begin{figure}[h]
  \centering
  \begin{tikzpicture}
    \def\h{1.6}
\def\d{-1}
\draw[thick, densely dotted] (-.5*\h, 0) to (4.5*\h,0) coordinate (e1);
\node[colorbox=blue] (q1) at (0,0) {$Q_1$};
\node[colorbox=orange] (q2) at (\h,0) {$Q_2$};
\node[colorbox=blue] (q3) at (2*\h,0) {$Q_3$};
\node[colorbox=orange] (q4) at (3*\h,0) {$Q_4$};
\node[colorbox=blue] (q5) at (4*\h,0) {$Q_5$};
\draw[conn=2pt, thick, ->] (q1) to ++(0,.7*\d) node[anchor=north] {$A_1$};
\draw[conn=2pt, thick, ->] (q2) to ++(0,.7*\d) node[anchor=north] {$A_2$};
\draw[conn=2pt, thick, ->] (q3) to ++(0,.7*\d) node[anchor=north] {$A_3\neq A_1$};
\draw[conn=2pt, thick, ->] (q4) to ++(0,.7*\d) node[anchor=north] {$A_4$};\draw[conn=2pt, thick, ->] (q5) to ++(0,.7*\d) node[anchor=north] {$A_5\neq A_3$};
\draw[conn=2pt, thin] (q1.north east) to[bend left] node[colorboxXS=cyan,above=2pt] {$\equiv$} (q3.north west);
\draw[conn=2pt, thin] (q3.north east) to[bend left] node[colorboxXS=cyan,above=2pt] {$\equiv$} (q5.north west);
   \end{tikzpicture}
  \caption{In a deterministic theory, binary questions with an interaction assumption lead to a contradiction:
   As between each pair of the equivalent questions,~$Q_1\equiv Q_3\equiv Q_5$, there is a non-implied question, the answers should all differ albeit~$\neg \neg A_1=A_5$.}
  \label{fig:disturbance_detTheory_2}
\end{figure}
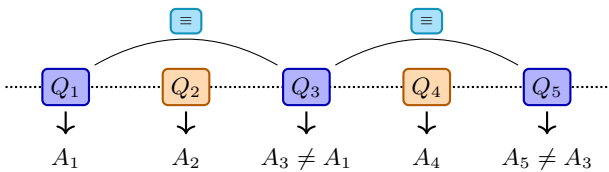

If we require the empirical evidence for an interaction to stem merely from inquiries about the system under consideration, we are lead to establish an interaction as a violation of the criterion for \emph{isolated systems}:
\emph{An interaction has to disturb the system.}
We are, thus, left with a \emph{contextual theory} in the sense of Paragraph~\ref{par:deterministic_theories}.
Let us consider successive inquiries about three questions that refer to the same system, i.e., that are $\sim^s$-equivalent, as depicted in Figure~\ref{fig:disturbance_detTheory}. 
Say we inquired about the first question~$Q_1$ and obtained an answer~$A_1$. 
Any inquiry about a non-implied question~$Q_2$---this implies that~$Q_2\not\equiv Q_1$---should alter the answer to questions~$Q_3\equiv Q_1$. 
In a deterministic theory with an interaction assumption, the answer~$A_3$ is then~$\neg A_1$ (``not~$A_1$'').
This, however, has problematic consequences for binary questions in~$\mathcal{Q}$:
Let us consider inquiries about $\equiv$-equivalent questions,~$Q_1$,~$Q_3$, and~$Q_5$. 
Between these, there are inquiries about non-implied questions,~$Q_2$ and~$Q_4$, as shown in Figure~\ref{fig:disturbance_detTheory_2}.
Both~$Q_2$ and~$Q_4$ disturb the answers to the other questions. 
Thus, inquiring about~$Q_5$ yields an answer 
\begin{equation*}
  A_5 = \neg A_3 = \neg (\neg A_1) = A_1\,.
\end{equation*}
The answers~$A_1=A_5$ to the questions~$Q_1\equiv Q_5$ let the system appear to be isolated---despite the inquiries of non-implied questions. 
By the pigeon-hole principle, this extends to any finite set of answers.
We arrive at an \emph{inconsistent triad}: A theory cannot 
\begin{enumerate*}[label={(\alph*)}]
  \item satisfy the interaction assumption, 
  \item refer to a finite number of questions, and
  \item be deterministic.
\end{enumerate*}
A ``spectator theory''~\cite{DeweyQFC} like classical mechanics~(see Section~\ref{sub:no_action_at_a_distance}) does not satisfy the first requirement.
Giving up the second requirement leads a theory based on, e.g., ternary logic as suggested by Reichenbach~\cite{Reichenbach1944,Putnam1957,Feyerabend1958,Kamlah1975}.
Loosening the last requirement, as we will do subsequently, takes us in the direction of quantum mechanics.

\paragraph{Probabilistic theories}\label{par:probabilistic_theories}
The inconsistent triad above can be escaped by turning to probabilistic theories with an adapted notion of isolated systems:
\emph{A system is \textbf{isolated} if and only if equivalent questions regarding the same system yield same answers with certainty.} 
That is, any question can be asked again with the same result:\footnote{With a suitable probability measure, a system is isolated if and only if~$\cProb{A_1=A_2}{Q_1\equiv Q_2} \ \forall Q_1,Q_2\in\mathcal{Q}$.}
\begin{IEEEeqnarray}{RL}\label{eqn:isol_prob}
  &S \ \text{is \emph{isolated} if and only if} \\
  &\forall Q\in\mathcal{Q} \ \text{with} \ (Q,A): \forall Q'\equiv Q: (Q', A)\, . \nonumber
\end{IEEEeqnarray}
Being isolated is not anymore a property of a single system, but rather of an equivalence class of such systems, e.g., across different runs of an experiment.
Thus, with the term ``system'' we refer from now on implicitly to such an equivalence class, and we assume that the set of questions~$\mathcal{Q}$ is associated with such a class of equivalent systems.
Then, a system is isolated if for arbitrarily many runs of the experiment, inquiring about $\equiv$-equivalent questions yields same answers---if in each run we can reassure ourselves by asking a $\equiv$-equivalent question about the identical system. 
Note that the answers \emph{across different runs} do not need to be the same---only within the same run.

\begin{figure}[h]
  \centering
  \begin{tikzpicture}
    \def\d{-0.35}
    \def\h{3.8}
    \node[colorbox=cyan] (isolSystem) at (0, \d) {isolated system};
\node[colorbox=orange] (equivQuest) at (0,-\d) {equivalent questions};
\draw[conn=4pt, ->] (isolSystem.east) .. controls ($(isolSystem.east)+(.8,0)$) and ($(equivQuest.east)+(.8,0)$) .. (equivQuest.east);
\draw[conn=4pt, <-] (isolSystem.west) .. controls ($(isolSystem.west)+(-.8,0)$) and ($(equivQuest.west)+(-.8,0)$) .. (equivQuest.west);
   \end{tikzpicture}
  \caption{Equivalent questions and isolated systems are merely meaningful concepts if they are considered together.}
  \label{fig:equivQuestions_isolSystems}
\end{figure}
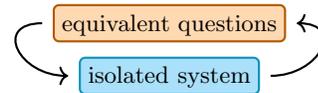

\paragraph{In retrospect}\label{par:in_retrospect}
The notion of an ``equivalence of questions'' has been fundamental for establishing a notion of ``isolated systems.''
The first should, however, not be regarded as logically prior to the latter---rather, the two have to be thought of as a mutually dependent, and to be made sense of together~(see Figure~\ref{fig:equivQuestions_isolSystems}).
In order to establish what is to be considered an equivalent question, one usually relies on an equivalent \emph{use}:
Two questions are then equivalent if they yield same answers in the \emph{same context}.

Similarly, the notion of ``identical systems'' is connected to the notion of isolated systems and the notion of equivalent questions.
The initial acceptance of means beyond quantum mechanics~(see Section~\ref{sec:meaning_interaction_sys}) offers the freedom to relate these notions without the need for an exhaustive reduction.

\subsection{Born rule in a contextual theory}\label{sub:born_rule_in_a_contextual_theory}
\noindent
The above considerations on the interaction assumption have left us with a \emph{contextual and probabilistic theory}:
For a given question, the theory yields a probability distribution over the answers, and this probability distribution depends on inquiries about other questions.
Generally, such a theory assigns to a finite set of questions~$\overline{Q}\subset\mathcal{Q}$ a joint probability distribution.
We assume that the possible answers to any questions in~$\mathcal{Q}$ are contained in a set~$\mathcal{A}$.
Further, we denote with~$\mathbb{P}'(\mathcal{Q})$ the set of \emph{finite ordered} subsets of~$\mathcal{Q}$.\footnote{The \emph{order} of measurements matters: 
A system can only be disturbed relative to \emph{prior} information about it. And this disturbance becomes only empirically tangible \emph{after} a corresponding interaction occurred.}
A probabilistic theory can then be regarded as a map that assigns to each ordered set~$\overline{Q}\in\mathbb{P}'(\mathcal{Q})$ a joint probability distribution,
\begin{equation}\label{eqn:born_cont_theory}
  T: \overline{Q} \mapsto  \Big\{ (x_1, \ldots, x_n) \in \mathcal{A}^n \mapsto P(x_1, \ldots, x_n) \Big\}\ . 
\end{equation}
The theory is contextual if the map $T$ does \emph{not} derive from a map
\begin{IEEEeqnarray}{RRCL}\label{eqn:born_noncont_theory}
  T':& \mathcal{Q} & \ \to \ & \text{Prob}(\mathcal{A}) \\ 
  & Q & \mapsto &  \big\{ x \in\mathcal{A} \mapsto P(x)\big\} \nonumber
\end{IEEEeqnarray}
that assigns to each question \emph{independently} a probability distribution.\footnote{This is not exactly the contextuality in the Kochen/Specker theorem~\cite{KS67,sep-kochen-specker}:
The contradiction in the Kochen/Specker theorem is merely dependent on the representation of observables as projectors on Hilbert spaces, the association of these observables with properties of a system, and the assumption that these observables have definite values independent of the experimental context.
The contradiction does \emph{not} involve the Born rule.
In our case, contextuality does not have to be a necessary consequence of the representation of measurements within the theory, but may as well be due to the form of the Born rule, i.e., how these operators are assigned probability distributions.}

\paragraph{Non-contextual theories}
In the case of a theory of the form~\eqref{eqn:born_noncont_theory}, the theory is equivalent to a relation~$\{(Q,x,P(x))\} \subset \mathcal{Q} \times \mathcal{A}\times [0,1]$ that associates each question-answer pair with a probability weight.
Such a theory might stem from a probability distribution over a Boolean lattice~$\mathcal{Q}\times\mathcal{A}$~(see Appendix~\ref{sec:qmlogics}).
Then, an interaction cannot ``disturb'' the measurement result, in the sense that it alters the probability distribution. 
The association of probability distributions is independent of other measurements.
An interaction is merely empirically tangible insofar as it \emph{changes} the probability distribution of other measurements in a contextual theory.
In other words, in a non-contextual theory we have no empirically tangible way to establish whether a system is isolated or not, in the sense above---whether or not an observer interacted with the system to inquire about a non-implied question~(as defined in Paragraph~\ref{par:transitivity}).

\paragraph{A general Born rule}
The map in \eqref{eqn:born_cont_theory} yields probability distributions over the possible answers of questions for a given set of questions---just as the Born rule does. 
It can thus be regarded as a generalization of the Born rule.
In the particular case of quantum mechanics, the states, according to Gleason's theorem~\cite{Gleason57}, correspond to probability distributions over an orthomodular, non-distributive lattice~(see Appendix~\ref{sec:qmlogics}).
Generally, the lattice of questions might be distributive: 
Then, the contextual character of the theory is rather a feature of the Born rule than a necessity of the formal representation of questions.

\subsection{Detecting an eavesdropper}\label{sub:general_bb84}
\noindent
Any theory satisfying the interaction assumption~\ref{IntAssum} allows to derive a key-exchange protocol as the one by Bennett and Brassard~(BB84)~\cite{BB84}.
If interactions can be traced empirically, then also the action of an eavesdropper provided that the he cannot guess what is an equivalent question: 
Alice inquires a system about a randomly chosen question~$Q_1\in\mathcal{Q}$, and then sends the system to Bob who also chooses a random question~$Q_2\in\mathcal{Q}$ and inquires about it.
If the two questions are equivalent,~$Q_1\equiv Q_2$, then Alice and Bob obtain the same answer.
For an eavesdropper to learn something about the obtained answers, he will have to inquire an equivalent question.
If he does not \emph{know} Alice's and Bob's question, he can merely \emph{guess} a question.
With non-zero probability, he will choose a non-implied question, and thus disturb the system. 
The disturbance reveals his intervention to Alice and Bob.

The above shows that the BB84 protocol relies crucially on the \emph{dichotomy between isolated systems and systems interacting with their environment}: 
Either a system is isolated, or it interacts with its environment. 
To learn something about the system, an interaction is necessary.
If interactions leave traces, then one can tell whether someone could have learnt something about that system.
Conversely, the BB84 protocol can be used to characterize what we mean by referring to ``isolated systems'' or to ``interactions:''
Instead of making interactions ``empirically tangible,'' we could equivalently have strived for making ``an eavesdropper's actions detectable.''
 \section{Interactions within an isolated~system}\label{sec:interactions_within_isolated_systems}
\noindent
So far, we have merely considered interactions of the environment (or of observers in the environment) \emph{with} the system under consideration.
We now turn to interactions \emph{within} an isolated system, i.e., between different parts of a joint system.
Two systems,~$S_1$ and~$S_2$, together can again be regarded as one system assuming the ability to refer to~$S_1$ and~$S_2$ suffices to refer to the corresponding combined system.
The joint system consisting of~$S_1$ and~$S_2$ is denoted~$S_1\times S_2$.

\subsection{Equivalent questions revisited}\label{sub:separable_questions}
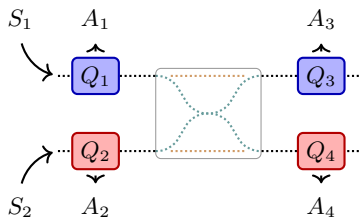
\begin{figure}[h]
  \begin{center}
    \begin{tikzpicture}
      \def\h{1.0}
\def\d{-.5}
\draw[thick, densely dotted, rounded corners=5pt] (-2*\h, 0) to (-.7*\h, 0);
\draw[thick, densely dotted, rounded corners=5pt] (2*\h, 0) to (.7*\h, 0);
\draw[thick, densely dotted, rounded corners=5pt] (-2*\h, 2*\d) to (-.7*\h, 2*\d);
\draw[thick, densely dotted, rounded corners=5pt] (2*\h, 2*\d) to (.7*\h, 2*\d);
\draw[thick, densely dotted, rounded corners=5pt, teal!70!black] (-.7*\h, 0) to (-.5*\h,0) to (-.2*\h, 1.0*\d) to (.2*\h, 1.0*\d) to (0.5*\h, 0) to (0.7*\h, 0);
\draw[thick, densely dotted, rounded corners=5pt, orange!70!black] (-.5*\h,0) to (0.5*\h, 0);
\draw[conn=2pt, <-] (-2*\h, 0) to[bend left] ++(-.5*\h, -\d) node[above] {$S_1$};
\draw[thick, densely dotted, rounded corners=5pt, teal!70!black] (-.7*\h, 2*\d) to (-.5*\h,2*\d) to (-.2*\h, 1.0*\d) to (.2*\h, 1.0*\d) to (0.5*\h, 2*\d) to (0.7*\h, 2*\d);
\draw[thick, densely dotted, rounded corners=5pt, orange!70!black] (-.5*\h,2*\d) to (0.5*\h, 2*\d);
\draw[conn=2pt, <-] (-2*\h, 2*\d) to[bend right] ++(-.5*\h, \d) node[below] {$S_2$};
\draw[rounded corners=2pt, fill=white, opacity=.4] (-.7*\h, -.2*\d) rectangle (.7*\h, 2.2*\d);
\node[colorbox=blue] (q1) at (-1.5*\h,0) {$Q_1$};
\node[colorbox=red] (q2) at (-1.5*\h,2*\d) {$Q_2$};
\node[colorbox=blue] (q3) at (1.5*\h,0) {$Q_3$};
\node[colorbox=red] (q4) at (1.5*\h,2*\d) {$Q_4$};
\draw[conn=2pt, thick, ->] (q1) to ++(0,-\d) node[anchor=south] {$A_1$};
\draw[conn=2pt, thick, ->] (q2) to ++(0,\d) node[anchor=north] {$A_2$};
\draw[conn=2pt, thick, ->] (q3) to ++(0,-\d) node[anchor=south] {$A_3$};
\draw[conn=2pt, thick, ->] (q4) to ++(0,\d) node[anchor=north] {$A_4$};

     \end{tikzpicture}
  \end{center}
  \caption{If two subsystems,~$S_1$ and~$S_2$, interact with each other after the respective inquiries about~$Q_1$ and~$Q_2$, then the answers obtained by inquiring about questions~$Q_3\equiv Q_1$ and~$Q_4\equiv Q_2$, differ from the respective previously obtained counterparts, as~$S_2$ is part of the environment of~$S_1$ and vice versa.
  If, however, we consider the pair~$(Q_1, Q_2)$ as a question about the combined system~$S_1\times S_2$, then we end up with an isolated system---the interaction is now \emph{within} the combined system---that yields \emph{different answers to equivalent questions}. 
  If we maintain that~$S_1\times S_2$ is isolated, then the interaction changes the notion of equivalent questions for~$S_1\times S_2$ with respect to the equivalence classes of~$S_1$ and~$S_2$.}
  \label{fig:sep_questions}
\end{figure}

\noindent
How does the notion of an isolated system for a combined system relate to the notion of isolated systems for the subsystems?
How do questions in~$\mathcal{Q}_{S_c}$ about the combined system~$S_c:=S_1\times S_2$ relate to questions about the subsystems, i.e., to elements in the Cartesian product~$\mathcal{Q}_{S_1}\times\mathcal{Q}_{S_2}$?
According to~\ref{IntAssum}, there is no difference if an observer interacted with~$S_1$ or another system~$S_2$ did. 
Let us consider the scenario depicted in Figure~\ref{fig:sep_questions}:
If one inquires about two equivalent questions~$Q_1\equiv Q_3$ about~$S_1$---before and after an interaction with the other subsystem~$S_2$---, then the answers must differ unless the interaction corresponds to inquiring about a question implied by~$Q_1$.
The same holds, vice versa, for~$S_2$ with respect to inquiries about equivalent questions~$Q_2\equiv Q_4$.
Thus, for a given interaction between~$S_1$ and~$S_2$, there are questions on the subsystem that detect the interaction.
The combined system~$S_c$, however, \emph{did not interact with its environment} between the pair of inquiries~$(Q_1,Q_2)$ and~$(Q_3,Q_4)$.
The combined system should still be isolated.
Therefore, \emph{if two subsystems interact with each other}, the notion of equivalent questions for~$S_c$ does not simply derive from the notion of equivalent questions for~$S_1$ and~$S_2$.
In particular, the pairs~$(Q_1,Q_2)$ and~$(Q_3,Q_4)$ are not equivalent questions for~$S_c$ unless the interaction corresponds to implied questions for both~$Q_1$ on~$S_1$ and~$Q_3$ on~$S_2$.

\subsection{The measurement problem}\label{sub:the_measurement_problem}
\noindent
If we assume that there is no qualitative difference between a system interacting with another, and a system interacting with an observer inquiring about a question, then one is tempted to regard an observer just \emph{as a system}.
As one inquired about a question for any system, one might inquire about the measurement result of an observer.
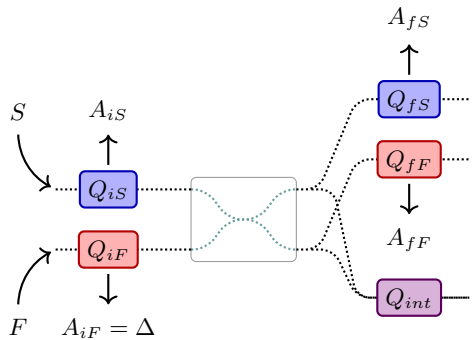
\begin{figure}[h]
  \centering
  \begin{tikzpicture}
    \def\d{-1.2}
    \def\h{3.8}
    \def\h{1.0}
\def\d{-.8}
\draw[thick, densely dotted, rounded corners=5pt] (-2.5*\h, 0) to (-.7*\h, 0);
\draw[thick, densely dotted, rounded corners=5pt] (.7*\h, 0) to (1.0*\h,0) to (1.5*\h,-1.5*\d) to (3*\h, -1.5*\d);
\draw[thick, densely dotted, rounded corners=5pt] (-2.5*\h, 1*\d) to (-.7*\h, 1*\d);
\draw[thick, densely dotted, rounded corners=5pt] (.7*\h, 1*\d) to (1.0*\h,1*\d) to (1.5*\h,-.5*\d) to (3*\h, -.5*\d) ;
\draw[thick, densely dotted, rounded corners=5pt] (.9*\h, 1*\d) to (1.2*\h,1*\d) to (1.5*\h,1.8*\d) to (3*\h, 1.8*\d) ;
\draw[thick, densely dotted, rounded corners=5pt] (.9*\h, 0*\d) to (1.2*\h,0*\d) to (1.5*\h,1.8*\d) to (3*\h, 1.8*\d) ;
\draw[thick, densely dotted, rounded corners=5pt, teal!70!black] (-.7*\h, 0) to (-.5*\h,0) to (-.2*\h, 0.5*\d) to (.2*\h, 0.5*\d) to (0.5*\h, 0) to (0.7*\h, 0);
\draw[thick, densely dotted, rounded corners=5pt, teal!70!black] (-.7*\h, 1*\d) to (-.5*\h,1*\d) to (-.2*\h, 0.5*\d) to (.2*\h, 0.5*\d) to (0.5*\h, 1.0*\d) to (0.7*\h, 1*\d);
\draw[conn=2pt, <-] (-2.5*\h, 0) to[bend left] ++(-.5*\h, -\d) node[above] {$S$};
\draw[conn=2pt, <-] (-2.5*\h, 1*\d) to[bend right] ++(-.5*\h, \d) node[below] {$F$};
\draw[rounded corners=2pt, fill=white, opacity=.4] (-.7*\h, -.2*\d) rectangle (.7*\h, 1.2*\d);
\node[colorbox=blue] (q1) at (-1.8*\h,0)   {$Q_{iS}$};
\node[colorbox=red] (q2) at (-1.8*\h,1*\d) {$Q_{iF}$};
\node[colorbox=blue] (q3) at (2.2*\h,-1.5*\d)    {$Q_{fS}$};
\node[colorbox=red] (q4) at (2.2*\h,-.5*\d)  {$Q_{fF}$};
\node[colorbox=violet] (q5) at (2.2*\h,1.8*\d)  {$Q_{int}$};
\draw[conn=2pt, thick, ->] (q1) to ++(0,-\d) node[anchor=south] {$A_{iS}$};
\draw[conn=2pt, thick, ->] (q2) to ++(0,\d) node[anchor=north]  {$A_{iF}=\Delta$};
\draw[conn=2pt, thick, ->] (q3) to ++(0,-\d) node[anchor=south] {$A_{fS}$};
\draw[conn=2pt, thick, ->] (q4) to ++(0,\d) node[anchor=north]  {$A_{fF}$};

   \end{tikzpicture}
  \caption{The behavior of the joint system~$S\times F$ is characterized by the dependence of the answers $A_{iS}$, $A_{iF}$, $A_{fS}$ and $A_{fF}$.
  If this joint system is isolated, then for \emph{any} pair of initial questions, there exists an equivalent question~$Q_{int}$ which yields the same answer. 
  The initial questions can then be chosen so that the system~$S$ gets disturbed.}
  \label{fig:gen_wigners_friend}
\end{figure}
This leads to a general Wigner's-friend experiment:
Let us assume that Wigner inquires a system~$F$ about an initial question $Q_{iF}$ with three possible answers, $\mathcal{A}_{iF} = \{0,1,\Delta\}$, and a system~$S$ about an initial question~$Q_{iS}$ with two possible answers,~$\mathcal{A}_{iS} = \{0,1\}$ as depicted in Figure~\ref{fig:gen_wigners_friend}.
The systems~$F$ and~$S$ then interact so that if Wigner initially obtains~$(A_{iS},\Delta)$, then he finally gets~$A_{fS} = A_{fF}=A_{iS}$ upon inquiring about~$Q_{fF}\equiv Q_{iF}$ and~$Q_{fS}\equiv Q_{iS}$.
This characterizes the joint, interacting system~$S\times F$:
The question~$(Q_{iS},Q_{iF})$ is equivalent to~$(Q_{fS},Q_{fF}')$ where~$Q_{fF}'$ is defined so that its answers are a suitable permutation of the answers of~$Q_{fF}$. 
In fact,~$(Q_{fS}, Q_{fF})$ is then implied by~$(Q_{fS},Q_{fF}')$ and inquiring about it does not disturb the joint system.

We now use that if $S\times F$ is isolated, then for \emph{any} pair of initial questions there exists an equivalent question~$Q_{int}$ that yields the same answer. 
Wigner changes the initial question~$Q_{iS}$ to another, non-equivalent~$Q_{iS}' \not\sim Q_{iS}$.
Also the combined question~$(Q_{iS}',Q_{iF})$ is in a different equivalence class,
\begin{equation*}
  Q_{int}'\equiv(Q_{iS}',Q_{iF}) \not\sim (Q_{iS},Q_{iF})\, .
\end{equation*}
Previously, the interaction corresponded for~$S$ to the inquiry about a question implied by~$Q_{iS}$: Inquiring about~$Q_{int}$ did not affect the answer~$A_{fS} = A_{iS}$.
This is not the case anymore for~$Q_{iS}'$: The inquiry about an question~$Q_{int}'\equiv (Q_{iS}', Q_{iF})$ potentially changes the answer to~$Q_{fS}$ and vice versa.
Either we inquire about~$(Q_{fS},Q_{fF})$ to obtain the result of the measurement carried out by Wigner's friend, associated with~$F$. 
Then, any subsequent measurement of $Q_{int}$ might yield a result that differs from the initial~$(A_{iS},A_{iF})$: We cannot regard~$S\times F$ as an isolated, interacting system.
Or we inquire about~$Q_{int}$ and we cannot be sure about the value the friend obtained in his measurement.\footnote{The correlation between the answers $A_{fS}$ and $A_{fF}$ might be preserved, as is the case in quantum mechanics. It is, however, not clear that this is generally the case.}

In quantum mechanics, the measurement problem revolves about the question whether a collapse can happen in an isolated system or not.
A collapse occurring inside an isolated system translates to the question, whether an \emph{isolated} system should behave \emph{as if} a question in the equivalence class of~$(Q_{iS},Q_{iF})$ had been inquired about.
This is at odds with the interaction assumption itself: Then the system would look as if it had been interacted with even though it did not.
 \section{Lattice of questions}
\label{sec:lattice_questions}
\noindent
In the binary case, questions in~$\mathcal{Q}$ can be regarded as statements that are either true or false.
A statement can imply another,
\begin{equation*}
  ( Q_1, \Bt) \Rightarrow ( Q_2, \Bt)\,,
\end{equation*}
where ``$\Rightarrow$'' is the notion of implication in ordinary language.
The implication yields an order relation~$\preceq$ on~$\mathcal{Q}$:
\begin{equation}\label{eqn:impl_q}
  Q_1 \preceq Q_2 \quad \text{if and only if} \quad (Q_1, \Bt) \Rightarrow (Q_2, \Bt)\,.
\end{equation}
If, in addition,~$(Q_2,\Bt)$ does \emph{not imply}~$(Q_1,\Bt)$, then we write~$Q_1\prec Q_2$.
The negation of a question $Q$ has an answer $\Bt$ if and only if the answer to $Q$ is $\Bf$,
\begin{equation*}
  ( \neg Q, \Bt) \Leftrightarrow ( Q, \Bf)\,.
\end{equation*}
We assume that there exists a question~$Q_t$ that is always answered with~$\Bt$---a \emph{tautological} question. 
Similarly we assume that there is a question~$\Bf$ that is always answered with~$\Bf$---the \emph{absurd} question.
Then, for all questions~$Q\in\mathcal{Q}$,~$Q_a\preceq Q \preceq Q_t$. Assuming that the joins and meets in the partially ordered set~$(\mathcal{Q}, \preceq)$ are unique, the set of questions together with the implication and negation forms a \emph{complemented lattice}~(for a brief, visual summary of lattices, see Appendix~\ref{sec:qmlogics}).
It seems natural to assume the negation to be involutional and order-reversing. 
This yields an \emph{orthocomplemented lattice}.

\subsection{Interlacing the lattice structure and the equivalence relation}\label{sub:interlacing}
\begin{figure}
  \centering
  \begin{tikzpicture}[thick,scale=1.2, every node/.style={transform shape}]
    \def\h{1}
\def\hshift{1.5}
\def\r{.5}
\def\rdot{1.2pt}
\fill[black] (0,2) circle (1.8pt) coordinate (o) node[font=\footnotesize, above] {1};
\fill[black] ($(o)+(0,-3.5*\h)$)  circle (1.8pt) coordinate (z) node[font=\footnotesize, below] {0};
\draw[thin, color=blue!70!black, densely dotted] (o) to ++(-\hshift,-1.*\h) coordinate (l1);
\draw[thin, color=blue!70!black, densely dotted] (z) to ++(-\hshift,.5*\h) ;
\draw[thin, color=blue!70!black, densely dotted] ($(l1)-(0,\h)$) ellipse[x radius=.7*\h, y radius=\h];
\begin{scope}[shift={($(l1)+(0,-1)$)}]
  \fill[black] ($(0,-.3)+(-150:\r)$) circle (\rdot) coordinate (a1) ;
  \fill[black] ($(0, .3)+(-210:\r)$) circle (\rdot) coordinate (b1) ;
  \fill[black] ($(0, .3)+(-330:\r)$) circle (\rdot) coordinate (g1) ;
  \fill[black] ($(0,-.3)+(-30:\r)$)  circle (\rdot) coordinate (d1) ;
  \fill[black] ($(0, .3)+(90:.1*\r)$)  circle (\rdot) coordinate (c11) ;
  \fill[black] ($(0,-.3)+(-90:.1*\r)$)  circle (\rdot) coordinate (c12) ;
\end{scope}
\draw[thin] (z) to (a1) (z) to (d1) (a1) to (b1) (b1) to (o) (o) to (g1) (g1) to (d1);
\draw[thin] (d1) to (c11) (a1) to (c11) (c11) to (o);
\draw[thin] (g1) to (c12) (b1) to (c12) (c12) to (z);
\draw[thin, color=violet!70!black, densely dotted] (o) to ++(0,-.75*\h) coordinate (l1);
\draw[thin, color=violet!70!black, densely dotted] (z) to ++(0,.75*\h) ;
\draw[thin, color=violet!70!black, densely dotted] ($(l1)-(0,\h)$) ellipse[x radius=.7*\h, y radius=\h];
\begin{scope}[shift={($(l1)+(0,-1)$)}]
  \fill[black] ($(0,-.3)+(-150:\r)$) circle (\rdot) coordinate (a2) ;
  \fill[black] ($(0, .3)+(-210:\r)$) circle (\rdot) coordinate (b2) ;
  \fill[black] ($(0, .3)+(-330:\r)$) circle (\rdot) coordinate (g2) ;
  \fill[black] ($(0,-.3)+(-30:\r)$)  circle (\rdot) coordinate (d2) ;
  \fill[black] ($(0, .3)+(90:.1*\r)$)  circle (\rdot) coordinate (c21) ;
  \fill[black] ($(0,-.3)+(-90:.1*\r)$)  circle (\rdot) coordinate (c22) ;
\end{scope}
\draw[thin] (z) to (a2) (z) to (d2) (a2) to (b2) (b2) to (o) (o) to (g2) (g2) to (d2);
\draw[thin] (d2) to (c21) (a2) to (c21) (c21) to (o);
\draw[thin] (g2) to (c22) (b2) to (c22) (c22) to (z);
\draw[thin, color=magenta!70!black, densely dotted] (o) to ++(\hshift,-.5*\h) coordinate (l1);
\draw[thin, color=magenta!70!black, densely dotted] (z) to ++(\hshift,1.*\h) ;
\draw[thin, color=magenta!70!black, densely dotted] ($(l1)-(0,\h)$) ellipse[x radius=.7*\h, y radius=\h];
\begin{scope}[shift={($(l1)+(0,-1)$)}]
  \fill[black] ($(0,-.3)+(-150:\r)$) circle (\rdot) coordinate (a3) ;
  \fill[black] ($(0, .3)+(-210:\r)$) circle (\rdot) coordinate (b3) ;
  \fill[black] ($(0, .3)+(-330:\r)$) circle (\rdot) coordinate (g3) ;
  \fill[black] ($(0,-.3)+(-30:\r)$)  circle (\rdot) coordinate (d3) ;
  \fill[black] ($(0, .3)+(90:.1*\r)$)  circle (\rdot) coordinate (c31) ;
  \fill[black] ($(0,-.3)+(-90:.1*\r)$)  circle (\rdot) coordinate (c32) ;
\end{scope}
\draw[thin] (z) to (a3) (z) to (d3) (a3) to (b3) (b3) to (o) (o) to (g3) (g3) to (d3);
\draw[thin] (d3) to (c31) (a3) to (c31) (c31) to (o);
\draw[thin] (g3) to (c32) (b3) to (c32) (c32) to (z);
\begin{pgfonlayer}{bg}
  \draw[thick, dashed, color=cyan!70!black] (a1) to (a2) to (a3);
  \draw[thick, dashed, color=cyan!70!black] (b1) to (b2) to (b3);
  \draw[thick, dashed, color=cyan!70!black] (g1) to (g2) to (g3);
  \draw[thick, dashed, color=cyan!70!black] (d1) to (d2) to (d3);
  \draw[thick, dashed, color=cyan!70!black] (c11) to (c21) to (c31);
  \draw[thick, dashed, color=cyan!70!black] (c12) to (c22) to (c32);
\end{pgfonlayer}
   \end{tikzpicture}
  \caption{The equivalence relation horizontally connects the vertical lattice structure.}
  \label{fig:lattice}
\end{figure}
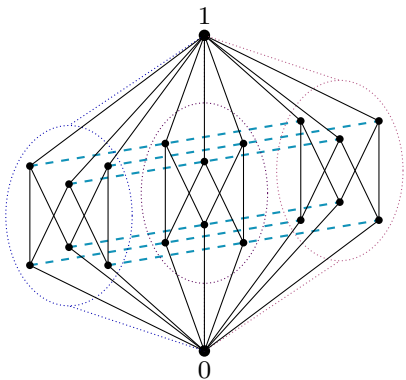

\noindent
It remains to join the lattice structure introduced above with the equivalence of questions required by verifiability.
We would like the lattice structure to be orthogonal to the $\sim$-equivalence in the following sense:
The lattice relates elements in~$\mathcal{Q}$ that correspond to the ``same time'' and the ``same run.'' 
The equivalence relation~$\sim$ relates questions ``horizontally'' across this ``vertical'' \mbox{(sub-)lattice} structure~(see Figure~\ref{fig:lattice}); questions corresponding to the ``same time'' and the ``same run'' are not $\sim$-equivalent.
Note that \emph{the $\sim$-equivalence of questions is then not equal to the bi-directional implication}
\begin{equation}\label{eqn:bi_dir_impl}
  ( Q_1, \Bt) \Rightarrow ( Q_2, \Bt)\ \text{and} \ ( Q_1, \Bt) \Leftarrow ( Q_2, \Bt)\,.
\end{equation}

\begin{figure}
  \centering
  \begin{tikzpicture}[thick,scale=1.2, every node/.style={transform shape}]
    \def\h{1}
\def\hshift{1.5}
\def\r{.5}
\def\rdot{1.2pt}
\fill[black] (0,2) circle (1.8pt) coordinate (o) node[font=\footnotesize, above] {1};
\fill[black] ($(o)+(0,-3.5*\h)$)  circle (1.8pt) coordinate (z) node[font=\footnotesize, below] {0};
\draw[thin, color=blue!70!black, densely dotted] (o) to ++(-\hshift,-1.*\h) coordinate (l1);
\draw[thin, color=blue!70!black, densely dotted] (z) to ++(-\hshift,.5*\h) ;
\draw[thin, color=blue!70!black, densely dotted] ($(l1)-(0,\h)$) ellipse[x radius=.7*\h, y radius=\h];
\begin{scope}[shift={($(l1)+(0,-1)$)}]
  \fill[black] ($(0,-.3)+(-150:\r)$) circle (\rdot) coordinate (a1) ;
  \fill[black] ($(0, .3)+(-210:\r)$) circle (\rdot) coordinate (b1) ;
  \fill[black] ($(0, .3)+(-330:\r)$) circle (\rdot) coordinate (g1) ;
  \fill[black] ($(0,-.3)+(-30:\r)$)  circle (\rdot) coordinate (d1) ;
  \fill[black] ($(0, .3)+(90:.1*\r)$)  circle (\rdot) coordinate (c11) ;
  \fill[black] ($(0,-.3)+(-90:.1*\r)$)  circle (\rdot) coordinate (c12) ;
\end{scope}
\draw[thin] (z) to (a1) (z) to (d1) (a1) to (b1) (b1) to (o) (o) to (g1) (g1) to (d1);
\draw[thin] (d1) to (c11) (a1) to (c11) (c11) to (o);
\draw[thin] (g1) to (c12) (b1) to (c12) (c12) to (z);
\draw[thin, color=violet!70!black, densely dotted] (o) to ++(0,-.75*\h) coordinate (l1);
\draw[thin, color=violet!70!black, densely dotted] (z) to ++(0,.75*\h) ;
\draw[thin, color=violet!70!black, densely dotted] ($(l1)-(0,\h)$) ellipse[x radius=.7*\h, y radius=\h];
\begin{scope}[shift={($(l1)+(0,-1)$)}]
  \fill[black] ($(0,-.3)+(-150:\r)$) circle (\rdot) coordinate (a2) ;
  \fill[black] ($(0, .3)+(-210:\r)$) circle (\rdot) coordinate (b2) ;
  \fill[black] ($(0, .3)+(-330:\r)$) circle (\rdot) coordinate (g2) ;
  \fill[black] ($(0,-.3)+(-30:\r)$)  circle (\rdot) coordinate (d2) ;
  \fill[black] ($(0, .3)+(90:.1*\r)$)  circle (\rdot) coordinate (c21) ;
  \fill[black] ($(0,-.3)+(-90:.1*\r)$)  circle (\rdot) coordinate (c22) ;
\end{scope}
\draw[thin] (z) to (a2) (z) to (d2) (a2) to (b2) (b2) to (o) (o) to (g2) (g2) to (d2);
\draw[thin] (d2) to (c21) (a2) to (c21) (c21) to (o);
\draw[thin] (g2) to (c22) (b2) to (c22) (c22) to (z);
\draw[thin, color=magenta!70!black, densely dotted] (o) to ++(\hshift,-.5*\h) coordinate (l1);
\draw[thin, color=magenta!70!black, densely dotted] (z) to ++(\hshift,1.*\h) ;
\draw[thin, color=magenta!70!black, densely dotted] ($(l1)-(0,\h)$) ellipse[x radius=.7*\h, y radius=\h];
\begin{scope}[shift={($(l1)+(0,-1)$)}]
  \fill[black] ($(0,-.3)+(-150:\r)$) circle (\rdot) coordinate (a3) ;
  \fill[black] ($(0, .3)+(-210:\r)$) circle (\rdot) coordinate (b3) ;
  \fill[black] ($(0, .3)+(-330:\r)$) circle (\rdot) coordinate (g3) ;
  \fill[black] ($(0,-.3)+(-30:\r)$)  circle (\rdot) coordinate (d3) ;
  \fill[black] ($(0, .3)+(90:.1*\r)$)  circle (\rdot) coordinate (c31) ;
  \fill[black] ($(0,-.3)+(-90:.1*\r)$)  circle (\rdot) coordinate (c32) ;
\end{scope}
\draw[thin] (z) to (a3) (z) to (d3) (a3) to (b3) (b3) to (o) (o) to (g3) (g3) to (d3);
\draw[thin] (d3) to (c31) (a3) to (c31) (c31) to (o);
\draw[thin] (g3) to (c32) (b3) to (c32) (c32) to (z);
\begin{pgfonlayer}{bg}
  \draw[thick, dashed, color=cyan!70!black] (a1) to (a2);
  \draw[thick, dashed, color=cyan!70!black] (c11) to (a2);
  \draw[thick, dashed, color=red!70!black] (a2) to (a3);
  \draw[thick, dashed, color=red!70!black] (d2) to (a3);
  \draw[thick, dashed, color=orange!70!black] (c21) to (c31);
  \draw[thick, dashed, color=orange!70!black] (c22) to (c31);
\end{pgfonlayer}
\draw[cyan!70!black, thick] (a1) circle (2.5pt) (c11) circle(2.5pt);
\draw[red!70!black, thick] (a2) circle (2.5pt) (d2) circle(2.5pt);
\draw[orange!70!black, thick] (c21) circle (2.5pt) (c22) circle(2.5pt);

   \end{tikzpicture}
  \caption{Joining equivalence classes as depicted here is problematic as it introduces empirically indistinguishable questions into the lattice structure.}
  \label{fig:lattice_join}
\end{figure}
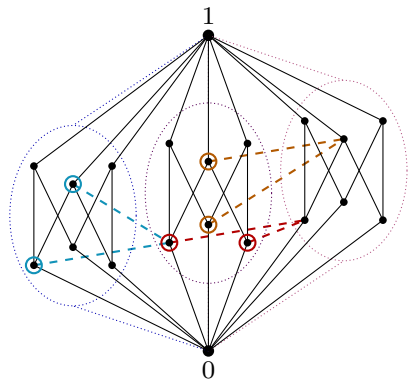

If there are two $\sim$-equivalent questions that correspond to the ``same time'' and the ``same run,''~$Q_1\sim Q_2, Q_1\neq Q_2$, then there are questions that are empirically not distinguishable: If the system is isolated they yield the same answer.
To distinguish the elements nonetheless, we have to introduce further characteristics for elements in~$\mathcal{Q}$ with some context in which they show empirically, thus loosing the uniformity of questions in~$\mathcal{Q}$.
If, then, any of the two is $\sim$-equivalent to questions at different times or in different runs, this corresponds, by transitivity, to joining equivalence classes as depicted in Figure~\ref{fig:lattice_join}.
This reminds of thermodynamics in two respects: There are different context, the micro- and the macro-context, in which questions are distinguishable or not; if equivalence classes join merely in \emph{one} direction, this yields an arrow of time, similar to the 2\textsuperscript{nd} Law. 
We refrain from a dualism of contexts here, and, thus, from allowing to join equivalence classes, as it counters the paradigm of empirical tangibility.

\begin{figure}
  \centering
  \begin{tikzpicture}[thick,scale=1.2, every node/.style={transform shape}]
    \def\h{1}
\def\hshift{1.5}
\def\r{.5}
\def\rdot{1.2pt}
\coordinate (o) at (0,2);
\coordinate (l1) at ($(o)+(-\hshift,-1.*\h)$);
\draw[thin, color=blue!70!black, densely dotted] ($(l1)-(0,\h)$) ellipse[x radius=.7*\h, y radius=\h];
\begin{scope}[shift={($(l1)+(0,-1)$)}]
  \fill[black] ($(0,+.3)+(90:\r)$) circle (1.8pt) coordinate (o1) node[font=\footnotesize, above] {1};
  \fill[black] ($(0,-.3)+(-90:\r)$) circle (1.8pt) coordinate (z1) node[font=\footnotesize, below] {0};
  \fill[black] ($(0,-.2)+(-150:\r)$) circle (\rdot) coordinate (a1) ;
  \fill[black] ($(0, .2)+(-210:\r)$) circle (\rdot) coordinate (b1) ;
  \fill[black] ($(0, .2)+(-330:\r)$) circle (\rdot) coordinate (g1) ;
  \fill[black] ($(0,-.2)+(-30:\r)$)  circle (\rdot) coordinate (d1) ;
  \fill[black] ($(0, .2)+(90:.1*\r)$)  circle (\rdot) coordinate (c11) ;
  \fill[black] ($(0,-.2)+(-90:.1*\r)$)  circle (\rdot) coordinate (c12) ;
\end{scope}
\draw[thin] (z1) to (a1) (z1) to (d1) (a1) to (b1) (b1) to (o1) (o1) to (g1) (g1) to (d1);
\draw[thin] (d1) to (c11) (a1) to (c11) (c11) to (o1);
\draw[thin] (g1) to (c12) (b1) to (c12) (c12) to (z1);
\coordinate (l1) at ($(o)+(0,-.75*\h)$);
\draw[thin, color=violet!70!black, densely dotted] ($(l1)-(0,\h)$) ellipse[x radius=.7*\h, y radius=\h];
\begin{scope}[shift={($(l1)+(0,-1)$)}]
  \fill[black] ($(0,+.3)+(90:\r)$) circle (1.8pt) coordinate (o2) node[font=\footnotesize, above] {1};
  \fill[black] ($(0,-.3)+(-90:\r)$) circle (1.8pt) coordinate (z2) node[font=\footnotesize, below] {0};
  \fill[black] ($(0,-.2)+(-150:\r)$) circle (\rdot) coordinate (a2) ;
  \fill[black] ($(0, .2)+(-210:\r)$) circle (\rdot) coordinate (b2) ;
  \fill[black] ($(0, .2)+(-330:\r)$) circle (\rdot) coordinate (g2) ;
  \fill[black] ($(0,-.2)+(-30:\r)$)  circle (\rdot) coordinate (d2) ;
  \fill[black] ($(0, .2)+(90:.1*\r)$)  circle (\rdot) coordinate (c21) ;
  \fill[black] ($(0,-.2)+(-90:.1*\r)$)  circle (\rdot) coordinate (c22) ;
\end{scope}
\draw[thin] (z2) to (a2) (z2) to (d2) (a2) to (b2) (b2) to (o2) (o2) to (g2) (g2) to (d2);
\draw[thin] (d2) to (c21) (a2) to (c21) (c21) to (o2);
\draw[thin] (g2) to (c22) (b2) to (c22) (c22) to (z2);
\coordinate (l1) at ($(o)+(\hshift,-.5*\h)$);
\draw[thin, color=magenta!70!black, densely dotted] ($(l1)-(0,\h)$) ellipse[x radius=.7*\h, y radius=\h];
\begin{scope}[shift={($(l1)+(0,-1)$)}]
  \fill[black] ($(0,+.3)+(90:\r)$) circle (1.8pt) coordinate (o3) node[font=\footnotesize, above] {1};
  \fill[black] ($(0,-.3)+(-90:\r)$) circle (1.8pt) coordinate (z3) node[font=\footnotesize, below] {0};
  \fill[black] ($(0,-.2)+(-150:\r)$) circle (\rdot) coordinate (a3) ;
  \fill[black] ($(0, .2)+(-210:\r)$) circle (\rdot) coordinate (b3) ;
  \fill[black] ($(0, .2)+(-330:\r)$) circle (\rdot) coordinate (g3) ;
  \fill[black] ($(0,-.2)+(-30:\r)$)  circle (\rdot) coordinate (d3) ;
  \fill[black] ($(0, .2)+(90:.1*\r)$)  circle (\rdot) coordinate (c31) ;
  \fill[black] ($(0,-.2)+(-90:.1*\r)$)  circle (\rdot) coordinate (c32) ;
\end{scope}
\draw[thin] (z3) to (a3) (z3) to (d3) (a3) to (b3) (b3) to (o3) (o3) to (g3) (g3) to (d3);
\draw[thin] (d3) to (c31) (a3) to (c31) (c31) to (o3);
\draw[thin] (g3) to (c32) (b3) to (c32) (c32) to (z3);
\begin{pgfonlayer}{bg}
  \draw[thick, dashed, color=cyan!70!black] (a1) to (a2) to (a3);
  \draw[thick, dashed, color=cyan!70!black] (b1) to (b2) to (b3);
  \draw[thick, dashed, color=cyan!70!black] (g1) to (g2) to (g3);
  \draw[thick, dashed, color=cyan!70!black] (d1) to (d2) to (d3);
  \draw[thick, dashed, color=cyan!70!black] (c11) to (c21) to (c31);
  \draw[thick, dashed, color=cyan!70!black] (c12) to (c22) to (c32);
  \draw[thick, dashed, color=cyan!70!black] (o1) to (o2) to (o3);
  \draw[thick, dashed, color=cyan!70!black] (z1) to (z2) to (z3);
\end{pgfonlayer}

   \end{tikzpicture}
  \caption{The equivalence relation horizontally connects partitions that form lattices.}
  \label{fig:partioned_lattices}
\end{figure}
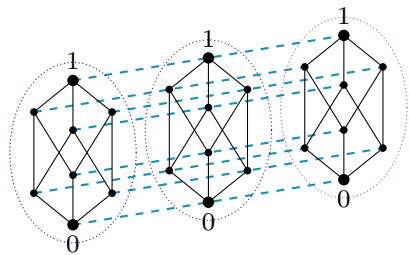

In the picture we have drawn this far (see Figure~\ref{fig:lattice}), the tautological and the absurd question reach across the horizontal structure: They bind together the sub-lattices.
To ensure that no two questions in the same sub-lattice are $\sim$-equivalent, we require that
\begin{equation}
  Q_1 \sim Q_2 \ \Rightarrow \ Q_1 \land Q_2 = 0 \ \text{and} \ Q_1\lor Q_2 = 1\,.
\end{equation}
This requirement is at odds with ortho-complementarity:
The complement is \emph{only} unique within a sub-lattice; not anymore across the entire lattice~$\mathcal{Q}$. 
Either we relax ortho-complementarity to hold merely for sub-lattices, or we assume that not the entire set~$\mathcal{Q}$ forms a lattice, but rather falls can be partitioned into ortho-complemented lattices and the $\sim$-equivalence reaches across these partitions as depicted in Figure~\ref{fig:partioned_lattices}.

\begin{figure}
  \centering
  \begin{tikzpicture}[thick,scale=1.2, every node/.style={transform shape}]
    \def\h{1}
\def\hshift{1.5}
\def\r{.5}
\def\rdot{1.2pt}
\coordinate (o) at (0,2);
\coordinate (l1) at ($(o)+(-\hshift,-1.*\h)$);
\draw[thin, color=blue!70!black, densely dotted] ($(l1)-(0,\h)$) ellipse[x radius=.7*\h, y radius=\h];
\begin{scope}[shift={($(l1)+(0,-1)$)}]
  \fill[black] ($(0,+.3)+(90:\r)$) circle (1.8pt) coordinate (o1);  \fill[black] ($(0,-.3)+(-90:\r)$) circle (1.8pt) coordinate (z1);  \fill[black] ($(0,-.2)+(-150:\r)$) circle (\rdot) coordinate (a1) ;
  \fill[black] ($(0, .2)+(-210:\r)$) circle (\rdot) coordinate (b1) ;
  \fill[black] ($(0, .2)+(-330:\r)$) circle (\rdot) coordinate (g1) ;
  \fill[black] ($(0,-.2)+(-30:\r)$)  circle (\rdot) coordinate (d1) ;
  \fill[black] ($(0, .2)+(90:.1*\r)$)  circle (\rdot) coordinate (c11) ;
  \fill[black] ($(0,-.2)+(-90:.1*\r)$)  circle (\rdot) coordinate (c12) ;
\end{scope}
\draw[thin, densely dotted] (z1) to (a1) (z1) to (d1) (a1) to (b1) (b1) to (o1) (o1) to (g1) (g1) to (d1);
\draw[thin, densely dotted] (d1) to (c11) (a1) to (c11) (c11) to (o1);
\draw[thin, densely dotted] (g1) to (c12) (b1) to (c12) (c12) to (z1);
\coordinate (l1) at ($(o)+(0,-.75*\h)$);
\draw[thin, color=violet!70!black, densely dotted] ($(l1)-(0,\h)$) ellipse[x radius=.7*\h, y radius=\h];
\begin{scope}[shift={($(l1)+(0,-1)$)}]
  \fill[black] ($(0,+.3)+(90:\r)$) circle (1.8pt) coordinate (o2) node[font=\footnotesize, above] {1};
  \fill[black] ($(0,-.3)+(-90:\r)$) circle (1.8pt) coordinate (z2) node[font=\footnotesize, below] {0};
  \fill[black] ($(0,-.2)+(-150:\r)$) circle (\rdot) coordinate (a2) ;
  \fill[black] ($(0, .2)+(-210:\r)$) circle (\rdot) coordinate (b2) ;
  \fill[black] ($(0, .2)+(-330:\r)$) circle (\rdot) coordinate (g2) ;
  \fill[black] ($(0,-.2)+(-30:\r)$)  circle (\rdot) coordinate (d2) ;
  \fill[black] ($(0, .2)+(90:.1*\r)$)  circle (\rdot) coordinate (c21) ;
  \fill[black] ($(0,-.2)+(-90:.1*\r)$)  circle (\rdot) coordinate (c22) ;
\end{scope}
\draw[thin] (z2) to (a2) (z2) to (d2) (a2) to (b2) (b2) to (o2) (o2) to (g2) (g2) to (d2);
\draw[thin] (d2) to (c21) (a2) to (c21) (c21) to (o2);
\draw[thin] (g2) to (c22) (b2) to (c22) (c22) to (z2);
\coordinate (l1) at ($(o)+(\hshift,-.5*\h)$);
\draw[thin, color=magenta!70!black, densely dotted] ($(l1)-(0,\h)$) ellipse[x radius=.7*\h, y radius=\h];
\begin{scope}[shift={($(l1)+(0,-1)$)}]
  \fill[black] ($(0,+.3)+(90:\r)$) circle (1.8pt) coordinate (o3);  \fill[black] ($(0,-.3)+(-90:\r)$) circle (1.8pt) coordinate (z3);  \fill[black] ($(0,-.2)+(-150:\r)$) circle (\rdot) coordinate (a3) ;
  \fill[black] ($(0, .2)+(-210:\r)$) circle (\rdot) coordinate (b3) ;
  \fill[black] ($(0, .2)+(-330:\r)$) circle (\rdot) coordinate (g3) ;
  \fill[black] ($(0,-.2)+(-30:\r)$)  circle (\rdot) coordinate (d3) ;
  \fill[black] ($(0, .2)+(90:.1*\r)$)  circle (\rdot) coordinate (c31) ;
  \fill[black] ($(0,-.2)+(-90:.1*\r)$)  circle (\rdot) coordinate (c32) ;
\end{scope}
\draw[thin, densely dotted] (z3) to (a3) (z3) to (d3) (a3) to (b3) (b3) to (o3) (o3) to (g3) (g3) to (d3);
\draw[thin, densely dotted] (d3) to (c31) (a3) to (c31) (c31) to (o3);
\draw[thin, densely dotted] (g3) to (c32) (b3) to (c32) (c32) to (z3);
\begin{pgfonlayer}{bg}
  \draw[thick, color=cyan!70!black] (a1) to (a2) to (a3);
  \draw[thick, color=cyan!70!black] (b1) to (b2) to (b3);
  \draw[thick, color=cyan!70!black] (g1) to (g2) to (g3);
  \draw[thick, color=cyan!70!black] (d1) to (d2) to (d3);
  \draw[thick, color=cyan!70!black] (c11) to (c21) to (c31);
  \draw[thick, color=cyan!70!black] (c12) to (c22) to (c32);
  \draw[thick, color=cyan!70!black] (o1) to (o2) to (o3);
  \draw[thick, color=cyan!70!black] (z1) to (z2) to (z3);
\end{pgfonlayer}

   \end{tikzpicture}
  \caption{If the $\sim$-equivalence relation preserves the lattice structure, then it can be be carried over to the equivalence classes.}
  \label{fig:lattice_equiv_classes}
\end{figure}
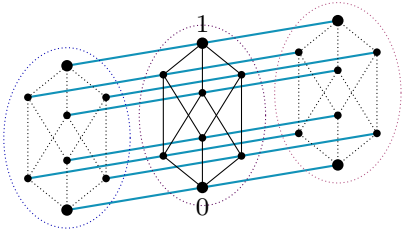

If we further assume that the equivalence relation \emph{preserves} the lattice structure by imposing
\begin{equation}
  Q_1 \sim Q_1' \ \text{and} \ Q_1 \preceq Q_2 \ \Rightarrow \ \exists! Q_2': \ Q_1' \preceq Q_2' \ \text{and} \ Q_2 \sim Q_2'\,,
\end{equation}
then we can lift the lattice structure to the $\sim$-equivalence classes as depicted in Figure~\ref{fig:lattice_equiv_classes}.

\subsection{Isolated systems for lattices of questions}\label{sub:isolated_systems_for_lattices_of_questions}
\noindent
One is drawn to assume that $\sim$-equivalence relation preserves the lattice structure by the following consistency requirement for the lattice structure and the notion of isolated systems:
\emph{Inquiring about implied questions does not disturb a system.}
Consider the scenario in Figure~\ref{fig:implied_questions}: 
If there exists a question~$Q_2'\in\mathcal{Q}$ such that
\begin{equation*}
  Q_1 \equiv Q_2' \ \text{and} \ Q_2' \preceq Q_2 \,.
\end{equation*}
Then inquiring about~$Q_2$ should not change the answer obtained when inquiring about~$Q_3\equiv Q_1$.
We have related the questions~$Q_1$ and~$Q_2$ following the lower path in the following diagram:
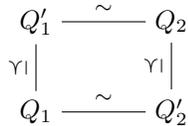
\begin{equation*}
  \begin{tikzpicture}
    \def\h{1.8}
    \def\u{1.2}
    \node[] (q1) at (0,0) {$Q_1$};
    \node[] (q2) at (\h,\u) {$Q_2$};
    \node[] (q1p) at (0,\u) {$Q_1'$};
    \node[] (q2p) at (\h,0) {$Q_2'$};
    \draw[] (q1) to node[midway, font=\footnotesize, rotate=90, above] {$\preceq$} (q1p) (q2) to node[midway, above, font=\footnotesize, rotate=90] {$\preceq$} (q2p);
    \draw[] (q1) to node[midway, font=\footnotesize, above] {$\sim$} (q2p) (q2) to node[midway, above, font=\footnotesize] {$\sim$} (q1p);
  \end{tikzpicture}
\end{equation*}
If the $\sim$-equivalence preserves the lattice structure, then diagram commutes, and there also exists a $Q_1'$ such that
\begin{equation*}
  Q_1 \preceq Q_1' \ \text{and} \ Q_1' \equiv Q_2 \,.
\end{equation*}
If there are multiple~$Q_1'$, then there will be joining equivalence classes. 
Thus, we demand here the uniqueness.
Similarly, we assume that inquiring about the negated question does not disturb the system.

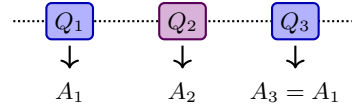
\begin{figure}[h]
  \centering
  \begin{tikzpicture}
    \def\h{1.5}
\def\d{-1}
\draw[thick, densely dotted] (-.5*\h, 0) to (2.5*\h,0) coordinate (e1);
\node[colorbox=blue] (q1) at (0,0) {$Q_1$};
\node[colorbox=violet] (q2) at (\h,0) {$Q_2$};
\node[colorbox=blue] (q3) at (2*\h,0) {$Q_3$};
\draw[conn=2pt, thick, ->] (q1) to ++(0,.7*\d) node[anchor=north] {$A_1$};
\draw[conn=2pt, thick, ->] (q2) to ++(0,.7*\d) node[anchor=north] {$A_2$};
\draw[conn=2pt, thick, ->] (q3) to ++(0,.7*\d) node[anchor=north] {$A_3=A_1$};
   \end{tikzpicture}
  \caption{Inquiring about an implied question should not disturb the answers}
  \label{fig:implied_questions}
\end{figure}

 \section{Conclusion}\label{sec:conclusion}
\noindent
If we require that there is ``no observation at a distance,'' and if we regard this requirement itself not exempt from our experience, then we are lead to assume that any interaction---including those that accompany our observations---leaves traces. 
Disturbance and complementarity in contextual, probabilistic theories with dichotomic notions of an \emph{isolated system} and an \emph{interaction} are then not a defect. 
Instead, they turn out to be a means to render interactions empirically tangible:
An interaction of a system with its environment alters the context, and, thus, also the probability distribution for other measurements performed on that system.
The necessity of a Born rule---a non-trivial map from ordered sets of measurements to probability distributions---is a consequence.

If interactions within isolated joint systems are qualitatively not different from interactions of systems with an observer, as they are regarded necessary by the interaction assumption, then there emerges a ``measurement problem:''
In a \emph{Gedankenexperiment} with encapsulated observers, \emph{\`{a} la} Wigner's friend, one can construct inquiries so that \emph{either} one can inquire about whether the encapsulated observer interacted with the system she measured \emph{or} what result the encapsulated observer obtained.
Either of the inquiries invalidates previously obtained answers to the respective other.
\emph{We cannot reconcile the idea of a measurement yielding a definite result with the measurement being an interaction within an isolated system.}
 \section{Epilogue}\label{sec:epilogue}
\noindent
\blockcquote[\S107 (own translation)\footnote{\blockcquote[\S107]{WittgPhiloUntersuchungen}[!]{Der Widerstreit [zwischen der tats\"{a}chlichen Sprache und unsrer Forderung nach der Kristallklarheit der Logik] wird unertr\"{a}glich; die Forderung droht nun zu etwas Leerem zu werden. --- Wir sind auf Glatteis geraten, wo die Reibung fehlt, also die Bedingungen in gewissem Sinne ideal sind, aber wir eben deshalb auch nicht gehen k\"{o}nnen. Wir wollen gehen; dann brauchen wir die \emph{Reibung}. Zur\"{u}ck auf den rauhen Boden}
 }]{WittgPhiloUntersuchungen}[!]{The antagonism [between the actual language and our demand for a crystal-clear logic] becomes unbearable; the demand is on the verge to become something empty. --- We got onto the clear ice where there is no friction, where the conditions are, in a sense, ideal, but also where we cannot walk. We want to walk; so, we need the friction. Back onto the rough ground}
The quest for certain, eternal, and unalterable knowledge has lead us onto the frictionless ice---the immutability of such knowledge necessitates a \emph{spectator theory}:
\blockcquote[]{DeweyQFC}[.]{[T]hat what is known is antecedent to the mental act of observation and inquiry, and is totally unaffected by these acts. [\ldots]
If the word `interaction' be used, it cannot denote that overt production of change it signifies in its ordinary and practical use. [\ldots]
The real object is the object so fixed in its regal aloofness that it is king to any beholding mind that may gaze upon it.
A spectator theory of knowledge is the inevitable outcome}
Conversely, rendering interactions, including those that go with our observations, subject to our experience---thereby restoring the ordinary meaning of an ``interaction''---requires to leave behind such spectator theories.
\emph{The measurement problem is the collision of a spectator notion of an observation with an ordinary idea of an interaction associated with the act of inquiry.}

\iffalse
The picture of an external cause behind sensory perceptions \emph{without the necessity of traceable interactions} may yield a crystal-clear logic---in the form of a non-contextual and deterministic theory.
Then, however, we find ourselves on the clear, friction-less ice: How can we meaningfully speak of an external cause if there is no empirical evidence of an interaction?

If, on the contrary, an interaction that is regarded necessary for an observation \emph{leaves empirically tangible traces}, one has the ground to meaningfully refer to an external cause. The measurement problem is here rather a characteristic of contextual, probabilistic theories than a defect of quantum mechanics.
\emph{The question is not necessarily how to \textbf{solve} the measurement problem but how it can be \textbf{made sense of}.}

  Knowledge [\ldots] is thought to be concerned with a region of being which is fixed in itself. [\ldots]
It can be approached through the medium of the apprehensions and demonstrations of thought, or by some other organ of mind, which does nothing to the real, except just to know it.
[\ldots]

\fi
 \begin{acknowledgments}
\noindent
  This work is supported by the Swiss National Science Foundation (SNF), the \emph{NCCR QSIT}, and the \emph{Hasler Foundation}. 
  We thank \"Amin Baumeler, Veronika Baumann, Cecilia Boschini, Xavier Coiteux-Roy, Paul Erker, Manuel Gil, and Alberto Montina for helpful discussions.
\end{acknowledgments}

\appendix
\section{Resolution restriction}\label{sub:resolution_restriction}
\noindent
In~\cite{Rovelli_relQM} and~\cite{Spekkens2007}, it is assumed that there is an upper bound on how much information one can have about a system. 
We examine how this \emph{resolution restriction} emerges in theories satisfying the interaction assumption.

\paragraph{Refinement}\label{par:refinement}
An ordered family of questions $\{Q_i\}_i$ that refer to the same system, i.e., that are $\sim^s$-equivalent, is called a \emph{refinement}, if any $Q_i$ is implied by its successor~$Q_{i+1}$ as defined in Paragraph~\ref{par:transitivity}, i.e.,
\begin{equation*}
  \forall i: \exists Q_i': Q_i'\equiv Q_{i+1} \land Q_i' \succ Q_{i+1}\,.
\end{equation*}

\paragraph{Reassurance}\label{par:reassurance}
A refinement can be \emph{reassured} in the following sense: 
If we inquire about the questions in its given order, then, at any time, we can reassure ourselves (i.e., inquire again about) previous questions,\footnote{Strictly speaking: we mean an inquiry about a $\equiv$-equivalent question.} without disturbing the system.
If the system is isolated, we will then obtain the same answer as to the equivalent question inquired about previously.

\iffalse
\begin{figure}[htpb]
  \centering
  \begin{tikzpicture}[scale=.65]
    \draw[very thick, densely dotted, ->, black!50] (.5,0) to (4.5, 0);
\foreach \i in {0,...,5} {
  \draw[ densely dotted, thick, blue!70!black] (\i,1.5) to (\i,-1.5) node[font=\footnotesize, black, below] {$Q_{\i}$};
  \draw[very thick, blue!70!black] (\i,1.2-.2*\i) to (\i,-1.2+.2*\i);
}
\node[font=\small] (M) at (2.5,3.0) {$M$};
\draw[shorten <=2pt, shorten >=2pt] (M.south) to[bend right] (1, 1.5);
\draw[shorten <=2pt, shorten >=2pt] (M.south) to[bend right] (2, 1.5);
\draw[shorten <=2pt, shorten >=2pt] (M.south) to[bend left] (3, 1.5);
\node[font=\footnotesize] (a0) at (-.8,.5) {$A_0$};
\draw[] (a0) to (0,.2);
\node[font=\footnotesize] (a5) at (5.8,.5) {$A_5$};
\draw[] (a5) to (5,.2);
   \end{tikzpicture}
  \caption{In a strict refinement, the answers are encapsulated subsets of $M$.}
  \label{fig:strict_refinement}
\end{figure}
\fi

\paragraph{Resolution restriction}\label{par:resolution_restriction}
If the lattice~$(\mathcal{Q}, \preceq)$ is atomic, then any refinement is finite.
Then there is only a finite number of questions that are not $\equiv$-equivalent and that can be inquired about without invalidating some of the questions.
The resolution restriction, i.e., the assumption that there is a maximal number of questions one can simultaneously know the answer to, can, thus, be regarded as a consequence of the interaction assumption together with the assumption that the lattice~$(\mathcal{Q}, \preceq)$ is atomic.

 \section{The ``state'' of a quantum ``system''}\label{sec:state_system}
\noindent
How does the above discussion change the perspective onto quantum mechanics?
In the following, we examine some repercussions.

\paragraph{Privileged questions}
We have argued in Section~\ref{sec:meaning_interaction_sys} that semantic intricacies taint the notion of an ultimate thing-in-itself, and, thus, also the notion of the \emph{state} exhaustively characterizing such an independently existing thing-in-itself---forming a ``state-in-itself.''
The interaction assumption adds to the scepticism regarding a reification of the state symbols: 
Even if the state-in-itself existed, the interaction assumption would bar the epistemic access to it.
After having asked a question~$Q$, one may ask refining questions. 
By the interaction assumption, however, there exists a question $Q'$ that allows to detect the inquiry of~$Q$.
Then, two pieces of information constituted from the inquiries about~$Q$ and~$Q'$ cannot be valid simultaneously.
Which of the two pieces of information does then tell us something about the state-in-itself?
Bohmian mechanics qualifies a ``position measurement'' as \emph{the} measurement that reveals the state-in-itself.
But there seems little in the way to single out the ``momentum measurement.''\footnote{Instead of assuming that ``everything is, in the end, a position measurement,'' one may follow the argumentation in Section~\ref{sec:meaning_interaction_sys} or return to Bohr's or Popper's argument~(see Section~\ref{sub:no_action_at_a_distance}) against a simple \emph{reading off} of pointer positions: 
If a measurement goes necessarily with an interaction, then everything might be regarded as a ``momentum measurement.''
In the ``reading-off-picture'' the position measurement is fundamental, in an ``interaction-picture'' the momentum measurement can be regarded as fundamental.}

\paragraph{Observer independence}
The formulation of the ``realism assumption'' in~\cite{PBR11} comes with similar issues:
\blockcquote[]{PBR11}[.]{One [assumption] is that a system \emph{has} a `real physical state'---not necessarily completely described by quantum theory, but objective and independent of the observer.
The assumption only needs to hold for systems that are isolated, and not entangled with other systems}
 The notion of an isolated system differs here from the one above:
The system-in-itself has a state-in-itself---independent of an observer.
This puts the state in a demon-like perspective\footnote{\blockcquote{PopperIndet1}{One might say that all these difficulties arise from the fact that the story of the Laplacean demon is an attempt to eliminate the vague and dangerous phrase `in principle.' 
For what it tries to explain is what we mean when we say that the future states of a system can be `in principle' predicted on the basis of a knowledge of past or present states.
`In principle' means here something like `not in practice, because human knowledge is never sufficiently precise and complete.' 
No wonder that, in attempting to explain what we mean by `in principle,' Laplace was driven to a superhuman intelligence. 
But the Laplacean demon is unsatisfactory, we may say, just because infinitely precise and complete knowledge is also `in principle' unattainable.}
 }:
If one can merely say something about that state after experience and corresponding interactions, then a state \emph{independent of the observer} remains a rather abstract concept.

\iffalse
\paragraph{The system-in-itself}\label{par:the_system_in_itself}
In Section~\ref{sub:the_notion_of_a_system}, it has been argued that the system is to be regarded as a \emph{context} for empirical inquiries.
The resolution restriction above adds to the scepticism whether a theory can establish reference to \emph{the same system}:
If that reference was established by validating characteristics of the system, then the resolution restriction restricts the number of non-equivalent characteristics one could inquire without disturbing the system.
Quantum mechanics is, in this sense, a contextual theory: 
The context sets what is considered a system.
\fi

\paragraph{Gleason's theorem} 
\label{par:Gleason}
Instead of regarding epistemically inaccessible symbols or answers to distinguished questions as the ``state-in-itself,'' we follow the direction of Gleason's theorem~\cite{Gleason57}:
If we choose to represent the~$\sim$-equivalence classes of questions by projectors on a Hilbert space, then---if the dimension is greater than two---the probability distributions over these projectors are in a one-to-one correspondence with the density matrices on that Hilbert space.
Therefore, the state symbols of quantum mechanics can be regarded as the probability distributions over the $\sim$-equivalence classes.
The \emph{collapse} is an initial choice of a probability distribution.
It corresponds to the assumption that, without further knowledge, we take a system to be isolated and, therefore, expect to obtain the same answer to subsequent, $\equiv$-equivalent questions.

\paragraph{Infinitely many questions}\label{par:infinitely_many_questions}
Even though one may assume that there are merely finitely many answers---we restrict ourselves to binary questions above---, this does not imply restrictions on the number of~$\sim$-equivalence classes of questions.
Then, the number of states must be equally unbounded.
This yields a variant of Hardy's theorem~\cite{HardyBaggage}.
In~\cite{JenningsLeifer16}, the continuity of the underlying mathematical structure is regarded as a strange aspect of quantum mechanics.\footnote{As noted in~\cite{JenningsLeifer16}, the step from infinite states to a continuous state space is shown in~\cite{Montina08}.}
\blockcquote[{\S}3.2]{JenningsLeifer16}[?]{Hardy's theorem together with results such as Holevo's bound and the discreteness of errors, show that precisely the opposite is the case: it is instead the continuity of quantum physics that is so strange. How can it be that we have a continuum of quantum states that ostenibly behave discretely but we do not have, and cannot have, an underlying discrete structure}
 The continuity merely appears strange if one elevates the quantum state to the thing-in-itself together with an assumption of epistemic transparency---the tenet of the immediate sensory accessibility of the thing-in-itself.

The conclusion that \blockcquote[{\S}3.2]{JenningsLeifer16}[]{even the most primitive quantum system must contain an infinite amount of information} relies on an imprecise use of the term ``information'': 
Even in an atomic lattice of questions with a resolution restrict resulting from the interaction assumption as argued in Appendix~\ref{sub:resolution_restriction}, there may be infinitely many $\sim$-equivalence classes of questions:
There might be only finite sets of questions so that all their answers can be reproduced without disturbing other questions in that set.
The ability of asking arbitrarily many questions should not be confused with the ability of having reproducible answers to all these questions.

 \section{Overview of quantum logic}
\label{sec:qmlogics}
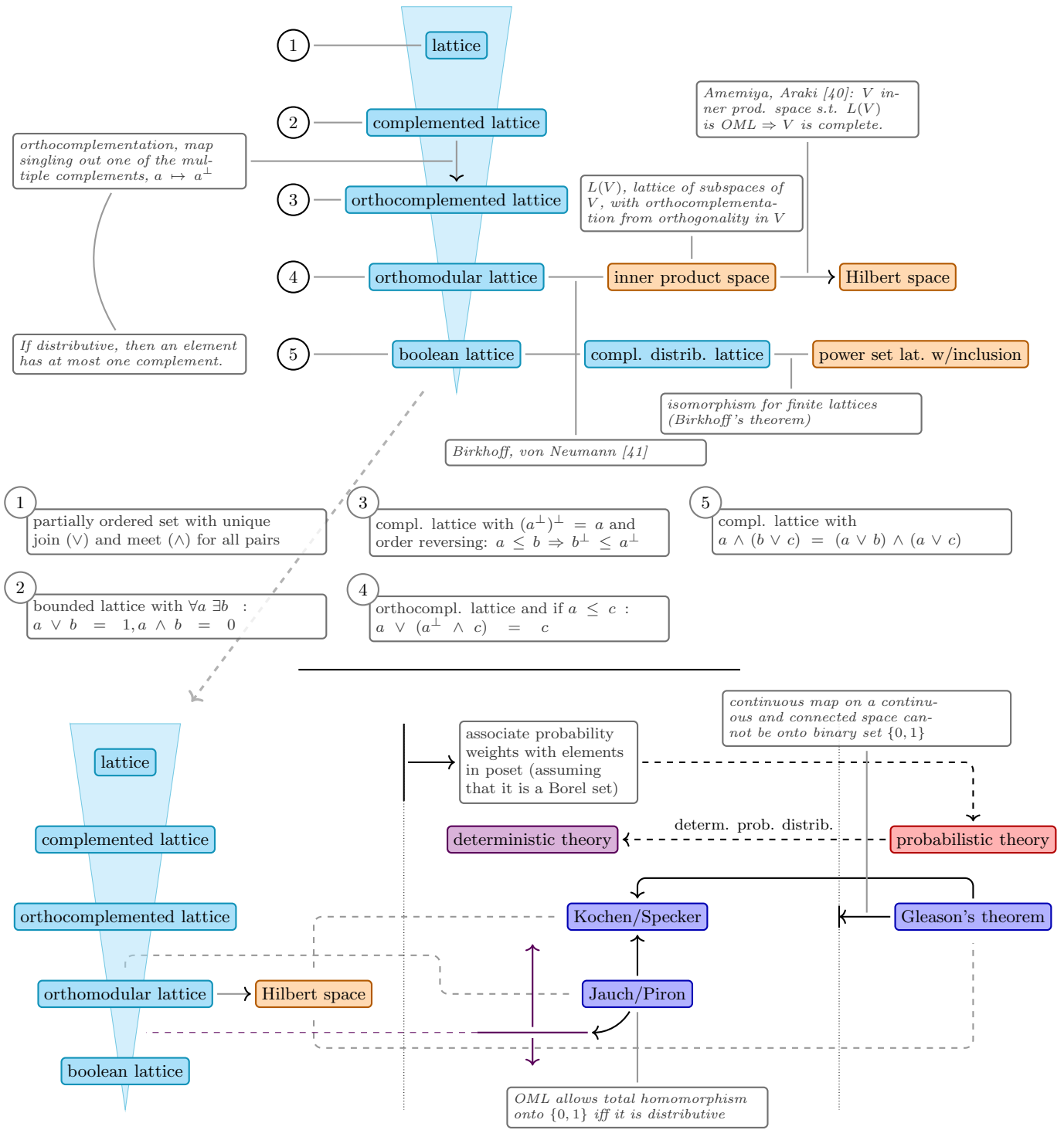
\begin{figure}[h]
  \centering
  \begin{tikzpicture}[quote/.append style={text width=5.2cm}, scale=.95, every node/.style={transform shape}]
    \def\d{-1.4}
    \def\h{3.8}
    \def\tw{5.8cm}
    \def\cLat{cyan}
\def\shl{-.7}
\def\cw{3.4cm}
\draw[cyan!80!black, fill=cyan!30, opacity=.5] (-1, -.5*\d) to (0,4.5*\d) to (1, -.5*\d) to cycle; 
\node[colorbox=\cLat] (lattice) at (0, 0) {lattice};
\draw[gconn=2pt] (lattice.west) to (\shl*\h, 0) node[draw, circle, anchor=east] {1};\node[colorbox=\cLat] (clattice) at (0, \d) {complemented lattice};
\draw[gconn=2pt] (clattice.west) to (\shl*\h, \d) node[draw, circle, anchor=east] {2};\node[comment=4cm, anchor=east] (oneElem) at (-1.0*\h, 4*\d) {If distributive, then an element has at most one complement.};
\node[colorbox=\cLat] (oclattice) at (0, 2*\d) {orthocomplemented lattice};
\draw[gconn=2pt] (oclattice.west) to (\shl*\h, 2*\d) node[draw, circle, anchor=east] {3};\draw[gconn=2pt, ->] (clattice.south) to coordinate[midway] (mw2) (oclattice.north);
\draw[gconn=2pt] (mw2) to ++(-1.0*\h, 0) node[comment=4cm, anchor=east] (orthcomplem) {orthocomplementation, map singling out one of the multiple complements, $a\mapsto a^{\perp}$};
\node[colorbox=\cLat] (omlattice) at (0, 3*\d) {orthomodular lattice};
\draw[gconn=2pt] (omlattice.west) to (\shl*\h, 3*\d) node[draw, circle, anchor=east] {4};\draw[gconn=2pt] (omlattice.east) to coordinate[midway] (oml1) ++(.3*\h, 0) node[colorbox=orange, anchor=west] (IPS) {inner product space};
\draw[gconn=2pt, ->] (IPS.east) to coordinate[midway] (mw4) ++(.3*\h, 0) node[colorbox=orange, anchor=west] (HilbertS) {Hilbert space};
\node[colorbox=\cLat] (boolattice) at (0, 4*\d) {boolean lattice};
\node[colorbox=\cLat, anchor=west]  (equivBL) at ($(boolattice.east)+(.3*\h, 0)$) {compl. distrib. lattice};
\draw[gconn=2pt] (equivBL.west) to (boolattice.east);
\draw[gconn=2pt] (boolattice.west) to coordinate[midway] (mw1) (\shl*\h, 4*\d)  node[draw, circle, anchor=east] {5};\node[colorbox=orange, anchor=west] (PSL) at ($(equivBL.east)+(.2*\h, 0)$) {power set lat.~w/inclusion};
\draw[gconn=2pt] (equivBL.east) to coordinate[midway] (mw3) (PSL.west);
\draw[gconn=2pt] (mw3) to ++(0,.5*\d) node[comment, anchor=north] {isomorphism for finite lattices\\ (Birkhoff's theorem)};
\node[quote=\cw, anchor=west] (l) at (-2.05*\h, 6.3*\d) {partially ordered set with unique join ($\lor$) and meet ($\land$) for all pairs};
\node[draw=gray, thick, circle, anchor=south east, fill=white] at ($(l.north west)+(.1,-.1)$) {1};
\node[quote=\cw, anchor=north west] (cl) at ($(l.south west)+(0, .5*\d)$) {bounded lattice with $\forall a \  \exists b: a\lor b= 1, a\land b=0$};
\node[draw=gray, thick, circle, anchor=south east, fill=white] at ($(cl.north west)+(.1,-.1)$) {2};
\node[quote=\cw, anchor=north west] (ocl) at ($(l.north east)+(.2*\h, .0*\d)$) {compl. lattice with $(a^{\perp})^{\perp}= a$ and order reversing: $a\leq b \Rightarrow b^{\perp} \leq a^{\perp}$};
\node[draw=gray, thick, circle, anchor=south east, fill=white] at ($(ocl.north west)+(.1,-.1)$) {3};
\node[quote=\cw, anchor=north west] (oml) at ($(ocl.south west)+(0, .5*\d)$) {orthocompl. lattice and if $a \leq c:$\\ $a\lor ( a^{\perp} \land c) = c$};
\draw[black, thick] ($(oml.south)+(-4,-.5)$) to ++(8,0);
\node[draw=gray, thick, circle, anchor=south east, fill=white] at ($(oml.north west)+(.1,-.1)$) {4};
\node[quote=\cw, anchor=north west] (bl) at ($(ocl.north east)+(.2*\h, .0*\d)$) {compl. lattice with\\ $a\land (b\lor c) = (a\lor b) \land (a\lor c)$};
\node[draw=gray, thick, circle, anchor=south east, fill=white] at ($(bl.north west)+(.1,-.1)$) {5};
\begin{pgfonlayer}{bg}
  \draw[gconn=2pt] (orthcomplem) to [bend right] (oneElem);
  \draw[gconn=2pt] (mw4) to ++(0, -1.8*\d) node[comment=3.8cm, anchor=south] {Amemiya, Araki~\cite{AA66}: $V$ inner prod. space s.t. $L(V)$ is OML $\Rightarrow$ $V$ is complete.};
  \draw[gconn=2pt] (IPS.north) to ++(0, -.4*\d) node[comment=3.8cm, anchor=south] {$L(V)$, lattice of subspaces of $V$, with orthocomplementation from orthogonality in $V$};
  \draw[gconn=2pt] (oml1) to ++(0,2.1*\d) node[comment, anchor=north] {Birkhoff, von Neumann~\cite{BirkhNeumLQM}};
\end{pgfonlayer}

     \begin{scope}[shift={(-6,-13)}]
      \def\cLat{cyan}
\draw[cyan!80!black, fill=cyan!30, opacity=.5] (-1, -.5*\d) to (0,4.5*\d) to (1, -.5*\d) to cycle; 
\node[colorbox=\cLat] (lattice1) at (0, 0) {lattice};
\node[colorbox=\cLat] (clattice1) at (0, \d) {complemented lattice};
\node[colorbox=\cLat] (oclattice1) at (0, 2*\d) {orthocomplemented lattice};
\node[colorbox=\cLat] (omlattice1) at (0, 3*\d) {orthomodular lattice};
\draw[gconn=2pt, ->] (omlattice1.east) to coordinate[midway] (mw41) ++(.2*\h, 0) node[colorbox=orange, anchor=west] (HilbertS1) {Hilbert space};
\node[colorbox=\cLat] (boolattice1) at (0, 4*\d) {boolean lattice};
\coordinate (c1) at ($(HilbertS1.east)+(.5*\h, .5*\d)$);
\coordinate (c1p) at ($(HilbertS1.east)+(.15*\h, .5*\d)$);
\draw[thick] (c1p |- {$(lattice1)+(0,.5*\d)$}) to coordinate[midway] (c3) ++(0, -\d);
\draw[conn=2pt, ->] (c3) to ++(1, 0) node[anchor=west, quote, text width=3.0cm] (theories) {associate probability weights with elements in poset (assuming that it is a Borel set)};
\draw[thick, violet!70!black] (c1) to coordinate[midway] (c2) ++(2, 0) coordinate (e1);
\draw[violet!70!black, dashed] (c1) to ++(-6,0);
\draw[thick, violet!70!black, ->] ($(c2)+(0, -.1)$) to ++(0, -.5);
\node[colorbox=violet] (detTh) at (c2 |- clattice1) {deterministic theory};
\draw[thick, violet!70!black, ->] ($(c2)+(0, .1)$) to coordinate[midway] (mw5) ++(0, 1.5);
\node[colorbox=red] (probTh) at ($(detTh)+(2.1*\h,0)$) {probabilistic theory};
\draw[conn=2pt, dashed, ->, rounded corners] (theories) to (theories -| probTh) to (probTh); 
\draw[conn=2pt, dashed, ->, rounded corners] (probTh) to node[above, font=\footnotesize] {determ. prob. distrib.} (detTh); 
\node[colorbox=blue] (gleason) at ($(probTh)+(0,\d)$) {Gleason's theorem};
\node[colorbox=blue] (KS) at ($(detTh)+(.5*\h,\d)$) {Kochen/Specker};
\node[colorbox=blue] (JP) at ($(KS)+(0,\d)$) {Jauch/Piron};
\draw[thick, shorten <=2pt, ->] (gleason.west) to coordinate[midway] (g2) ++(-1,0) coordinate (g1);
\draw[densely dotted] (c1p |- {$(lattice1)+(0,-.5*\d)$}) to (c1p |- {$(boolattice1)+(0,.5*\d)$});
\draw[densely dotted] (g1 |- {$(lattice1)+(0,-.5*\d)$}) to (g1 |- {$(boolattice1)+(0,.5*\d)$});
\draw[thick] ($(g1)+(0,.2)$) to ++(0,-.4);
\draw[gconn=2pt] (g2) to ++(0, -2.2*\d) node[comment, text width=5cm, anchor=south] {continuous map on a continuous and connected space cannot be onto binary set $\{0,1\}$};

\begin{pgfonlayer}{bg}
  \draw[gconn=6pt, dashed, rounded corners] (HilbertS1.south) to (HilbertS1 |- {(0,3.7*\d)}) to ({(0,3.7*\d)} -| gleason) to (gleason);
  \draw[gconn=6pt, dashed, rounded corners] (HilbertS1) to (HilbertS1 |- KS) to (KS);
  \draw[conn=2pt, ->, rounded corners] (gleason) to (gleason |- {$(gleason)+(0,-.5*\d)$}) to (KS |- {$(KS)+(0,-.5*\d)$}) to (KS);
  \draw[conn=2pt, ->] (JP) to (KS);
  \draw[gconn=6pt, rounded corners, dashed] (omlattice1) to ++(0,-.5*\d) to ++(1.5*\h, 0) to ++(0,.5*\d) to (JP);
  \draw[gconn=2pt] (JP) to ++(0,1.2*\d) node[comment, anchor=north] {OML allows total homomorphism onto $\{0,1\}$ iff it is distributive};   \draw[conn=2pt, ->] (JP) to[bend left] (e1);
\end{pgfonlayer}
     \end{scope}
    \begin{pgfonlayer}{bg}
      \draw[very thick, dashed, gray!60, rounded corners, ->] ($(boolattice.south)+(-.6,.3*\d)$) to ($(lattice1.north)+(1.2,-.6*\d)$);
    \end{pgfonlayer}
  \end{tikzpicture}
  \caption{A brief summary of lattices and quantum logic.}
  \label{fig:qm_logics}
\end{figure}

\noindent
Figure~\ref{fig:qm_logics} gives a brief introduction to lattices and summarizes import result in quantum logic (see~\cite{sep-qmlogics,Coecke2000,GG1975,bacciagaluppi2009}).
The terms in the blue boxes form a sequence of narrowing definitions, starting from a lattice (Box~\raisebox{.5pt}{\textcircled{\raisebox{-.9pt} {1}}}):
A partially ordered set~$(L,\leq)$ is called a lattice if any two elements~$a,b\in L$ have a unique least upper bound or \emph{join},~$a \lor b$, and a unique greatest lower bound or \emph{meet},~$a\land b$. 
A lattice is \emph{bounded} if there exists a greatest element,~$1\geq a \ \forall a \in L$, and a least element,~$0\leq a \ \forall a\in L$.
A bounded lattice is \emph{complemented} if for all~$a\in L$ there exists a~$b\in L$ so that~$a\land b = 0$ and~$a\lor b=1$ (Box~\raisebox{.5pt}{\textcircled{\raisebox{-.9pt} {2}}}).
If one can choose among the complements of elements~$a\in L$ one, such that $(a^{\perp})^{\perp}=a$, and the complement reverses the order, i.e.,~$a\leq b \Rightarrow b^{\perp} \leq a^{\perp}$, then such a map is called an \emph{orthocomplementation} and the resulting lattice is called \emph{orthocomplemented} (Box~\raisebox{.5pt}{\textcircled{\raisebox{-.9pt} {3}}}).
In the lattice of closed subspaces of a Hilbert space ordered by the set-inclusion of their ranges,~$L(\mathcal{H})$, orthogonality yields an orthocomplementation. 
The resulting lattice is \emph{orthomodular}, i.e., $\forall a,b\in L, a \leq b: a\lor ( a^{\perp} \land b) = b$ (Box~\raisebox{.5pt}{\textcircled{\raisebox{-.95pt} {4}}}). 
The lattice~$L(\mathcal{H})$ is, however, \emph{not distributive}, and, thus, not a Boolean lattice (Box~\raisebox{.5pt}{\textcircled{\raisebox{-.95pt} {5}}}):
The theorems by Jauch and Piron~\cite{JP1963}, Kochen and Specker~\cite{KS67}, and Gleason~\cite{Gleason57} limit the possibility of quantum mechanics being a non-contextual theory. In the contrary, we develop a perspective in which a theory \emph{should be contextual}, and, thus, not require the structure of a distributive lattice.

\onecolumngrid
 
\end{document}